\documentclass[11pt]{article}

\usepackage{epsfig,graphics}

\usepackage{amssymb}

\newcommand{\bPi}{\mbox{\boldmath $\Pi$}}
\newcommand{\bPsi}{\mbox{\boldmath $\Psi$}}
\newcommand{\bP}{\mbox{\boldmath $P$}}

\newcommand{\beqn}{\begin{eqnarray}}
\newcommand{\eeqn}{\end{eqnarray}}
\newcommand{\be}{\begin{equation}}
\newcommand{\ee}{\end{equation}}
\newcommand{\ba}{\begin{array}}
\newcommand{\ea}{\end{array}}
\newcommand{\R}{{\rm\bf R}}
\newcommand{\C}{{\rm\bf C}}
\newcommand{\pa}{\partial}

\newcommand{\re}{\ref}
\newcommand{\ci}{\cite}
\newcommand{\la}{\label}
\newcommand{\bfr}{\begin{flushright}}
\newcommand{\efr}{\end{flushright}}
\newcommand{\bfl}{\begin{flushleft}}
\newcommand{\efl}{\end{flushleft}}
\newcommand{\fr}{\frac}
\newcommand{\ov}{\overline}
\newcommand{\ti}{\tilde}
\newcommand{\st}{\stackrel}

\newcommand{\si}{\sigma} 
\newcommand{\al}{\alpha}

\newcommand{\ds}{\displaystyle}
\newcommand{\ts}{\textstyle}

\newcommand{\cE}{{\cal E}} \newcommand{\cF}{{\cal F}}
\newcommand{\cH}{{\cal H}}

\newcommand{\cO}{{\cal O}}
\newcommand{\cS}{{\cal S}}

\newcommand{\bI}{{\bf I}}
\newcommand{\cT}{{\cal T}}
\newcommand{\cZ}{{\cal Z}}

\newcommand{\5}{{\hspace{0.5mm}}}
\newcommand{\ve}{\varepsilon}

\newcommand{\we}{\wedge}
\newcommand{\de}{\delta}\newcommand{\De}{\Delta}

 \newcommand{\ga}{\gamma}
\newcommand{\om}{\omega}
\newcommand{\Om}{\Omega}

\newcommand{\na}{\nabla}

\newcommand{\lam}{\lambda} \newcommand{\ka}{\varkappa}
\newcommand{\Lam}{\Lambda}
\newcommand{\co}{{\rm const}}
\newcommand{\supp}{{\rm supp\5}}

\newcommand{\Br}{|\kern-.25em|\kern-.25em|}
\newcommand{\brr}{{|\kern-.15em|\kern-.15em|\kern-.15em}\,}
\newcommand{\ddd}{\st{.\kern-.07em.\kern-.07em.}}
\def\N{{\rm I\kern-.1567em N}}                              
\def\R{{\rm I\kern-.1567em R}}                              
\def\C{{\rm C\kern-4.7pt                                    
\vrule height 7.7pt width 0.4pt depth -0.5pt \phantom {.}}}
\def\Z  {{\sf Z\kern-4.5pt Z}}                                
\def\Re {{\rm Re\, }}                                       
\def\Im {{\rm Im\,}}                                        

\begin{document}

\renewcommand{\theequation}{\thesection.\arabic{equation}}
\newtheorem{theorem}{Theorem}[section]
\renewcommand{\thetheorem}{\arabic{section}.\arabic{theorem}}
\newtheorem{definition}[theorem]{Definition}
\newtheorem{deflem}[theorem]{Definition and Lemma}
\newtheorem{lemma}[theorem]{Lemma}
\newtheorem{example}[theorem]{Example}
\newtheorem{remark}[theorem]{Remark}
\newtheorem{remarks}[theorem]{Remarks}
\newtheorem{cor}[theorem]{Corollary}
\newtheorem{pro}[theorem]{Proposition}

\newcommand{\bd}{\begin{definition}}
 \newcommand{\ed}{\end{definition}}
\newcommand{\bt}{\begin{theorem}}
 \newcommand{\et}{\end{theorem}}
\newcommand{\bqt}{\begin{qtheorem}}
 \newcommand{\eqt}{\end{qtheorem}}

\newcommand{\bp}{\begin{pro}}
 \newcommand{\ep}{\end{pro}}

\newcommand{\bl}{\begin{lemma}}
 \newcommand{\el}{\end{lemma}}
\newcommand{\bc}{\begin{cor}}
 \newcommand{\ec}{\end{cor}}

\newcommand{\bex}{\begin{example}}
 \newcommand{\eex}{\end{example}}
\newcommand{\bexs}{\begin{examples}}
 \newcommand{\eexs}{\end{examples}}

\newcommand{\bexe}{\begin{exercice}}
 \newcommand{\eexe}{\end{exercice}}

\newcommand{\br}{\begin{remark} }
 \newcommand{\er}{\end{remark}}
\newcommand{\brs}{\begin{remarks}}
 \newcommand{\ers}{\end{remarks}}

\newcommand{\ft}{\footnote}

\newcommand{\pru}{{\bf Proof~~}}


\phantom{}

\vspace{1cm}

\begin{center}

\bigskip

\bigskip
{\huge\bf
Scattering of Solitons for Coupled

\bigskip

Wave-Particle Equations
}

\vspace{1cm}

{\bf Running head: } Soliton Scattering for Wave-Particle Eqs

\vspace{1cm} {\large Valery Imaykin}\,\footnote{Supported partially
by Austrian Science Foundation (FWF), by
research grants of DFG and RFBR.}
\medskip\\
{\it Faculty of Mathematics\\
TU Munich,
Garching, 85747 Germany}

\vspace{1cm}

{\large Alexander Komech}$^1\,$\,\footnote{Supported by the Alexander von Humboldt Research Award.}
\medskip\\
{\it Faculty of Mathematics of
Vienna University and\\
Institute for Information Transmission Problems RAS.}

\vspace{1cm}

{\large Boris Vainberg}\,\footnote{Supported partially by NSF grant DMS-0706928.}\medskip\\
{\it Department of Mathematics and Statistics,\\ UNC at Charlotte,
Charlotte, NC 28223, USA}

\end{center}

\vspace{2cm}

\begin{abstract}

We establish a long time soliton asymptotics for a nonlinear
system of wave equation coupled to a charged particle. The coupled
system has a six dimensional manifold of soliton solutions. We
show that in the large time approximation, any solution, with an
initial state close to the solitary manifold, is a sum of a
soliton and a dispersive wave which is a solution to the free wave
equation. It is assumed that the charge density satisfies
Wiener condition which is a version of Fermi Golden Rule, and
that the momenta of the charge distribution vanish up to the fourth order. The proof is
based on a development of the general strategy introduced by
Buslaev and Perelman:
symplectic projection in Hilbert space onto the solitary manifold,
modulation equations for the parameters of the projection, and
decay of the transversal component.
\end{abstract}

\vspace{1.5cm}



\newpage


\setcounter{equation}{0}

\section{Introduction}

Our paper concerns an old mathematical problem of nonlinear
field-particle interaction. A charged particle radiates a field
which acts on the particle etc. This self-action is probably
responsible for some crucial features of the process: asymptotically
 uniform motion of the particle, increment of the particle's mass etc.
(see \ci{Sp}, Part I). The problem has many different appearances: for a classical
particle coupled to a scalar or Maxwell field, for coupled Maxwell-Schr\"odinger
 or Maxwell-Dirac equations, for the corresponding second-quantized
 equations etc.

One of the main goals of mathematical investigation of this problem
 is studying soliton-type long time asymptotics and
asymptotic stability of soliton solutions.
The first results in this direction have been discovered for KdV
equation and other {\it complete integrable equations}. For KdV
equation, any solution with sufficiently smooth and rapidly
decaying initial data converges to a finite sum of solitons moving
to the right, and a dispersive wave moving to the left. A complete
survey and proofs can be found in \ci{EH81,FZ93}.

For non-integrable equations, the long time convergence of the
solution to a soliton part and dispersive wave was obtained first
by Soffer and Weinstein in the context of
 $U(1)$-invariant  Schr\"odinger equation with potential, \ci{SW88,SW90,SW92}.
The extension to translation invariant equations was obtained by
 Buslaev and Perelman \ci{BP1,BP2,BP3,BS} for 1D Schr\"odinger equation,
and by Miller, Pego and Weinstein for 1D modified KdV and RLW equations,
\ci{MW96,PW92,PW94}. Later the results were developed by
Bambusi and Cuccagna \ci{BC08,Cu, Cu03,Cu10},
and
Martel
and Merle \ci{MM08}.

In \ci{BP1,BP2,BP3,BS} the long time convergence is obtained for
translation invariant and $U(1)$-invariant nonlinear Schr\"odinger
equation: for any finite-energy solution $\psi(x,t)$ with initial
data close to a
 soliton $\psi_{v_0}(x-v_0t-a_0)e^{i\om_0t}$,
the following asymptotics hold: \be\la{Bus}
\psi(x,t)=\psi_{v_\pm}(x-v_\pm t-a_\pm)e^{i\om_\pm
t}+W_0(t)\psi_{\pm}+ r_\pm(x,t),\,\,\,t\to\pm\infty. \ee Here the
first term of the right hand side is a soliton with parameters
$v_\pm$, $a_\pm$, $\om_\pm$ close to $v_0$, $a_0$, $\om_0$, the
function  $W_0(t)\psi_{\pm}$ is a dispersive wave which is a
solution to the free Schr\"odinger equation,
 and the remainder $r_\pm(x,t)$ converges to zero in the global $L^2$-norm.

In the present paper we consider a scalar real-valued wave field $\psi (x)$ in $\R^3$,
coupled to a relativistic particle with position $q$ and momentum $p$,
governed by
\beqn\la{system0}
\ba{lll}
\dot{\psi}(x,t)=\pi (x,t), &
\dot{\pi}(x,t)=\Delta \psi (x,t)
-\rho (x-q(t)), & x\in\R^3,
\\
~ &  \\
\dot{q}(t)=p(t)/\sqrt{1+p^2}, & \dot{p}(t)=\displaystyle\int
\psi
(x,t)\,\nabla \rho (x-q(t))dx. &
\ea
\eeqn
This is a Hamilton system with the Hamilton functional
\be\la{hamilq0}
{\cal H}(\psi ,\pi ,q,p)=\frac 12\int\Big(|\pi(x)|^2+
|\nabla \psi (x)|^2\Big)dx+\int \psi (x)\rho (x-q)dx+
\sqrt{1+p^2}.
\ee
The first two equations for the fields are equivalent to the
 wave equation with the source $\rho(x-q)$.
We have set the mechanical mass of the particle and the speed
of wave propagation equal to one.
The case of the point particle corresponds to $\rho (x)=\de(x)$ and then
the interaction term in the Hamiltonian is simply
$\psi (q)$.
However, in this case the Hamiltonian is unbounded from below which
leads to the ill-posedness of the problem,
that is also known as ultraviolet divergence.
Therefore we smooth the coupling by the function $\rho (x)$
following the
``extended electron''
strategy proposed by M.~Abraham for the Maxwell field.
 In analogy to the Maxwell-Lorentz equations we call $\rho$ the
``charge distribution''.
Finally, the form of the last two equations in
(\re{system0})
is determined by the choice of the relativistic
kinetic energy $ \sqrt{1+p^2}$ in (\re{hamilq0}).

Let us write the system (\re{system0}) as
\be\la{HPDE} \dot
Y(t)=F(Y(t)),~~~~~t\in\R, \ee where
$Y(t):=(\psi(x,t),\pi(x,t),q(t),p(t))$. The system
(\re{system0}) is translation-invariant and admits soliton
solutions \be\la{sosol} Y_{a,v}(t)=(\psi_v(x-vt-a),\pi_v(x-vt-a),
vt+a,p_v), ~~~~~~~p_v=v/{\sqrt{1-v^2}} \ee for all $a,v\in\R^3$
with $|v|<1$ (see (\re{stfch}) to (\re{sol}) below). The states
$S_{a,v}:=Y_{a,v}(0)$ form the solitary manifold
\be\la{soman}
{\cal S}:=\{ S_{a,v}: a,v\in\R^3, |v|<1 \}.
\ee
Our main result (announced in \ci{IKV06}) is
the soliton asymptotics of type (\re{Bus}),
\be\la{Si}
(\psi(x,t), \pi(x,t))\sim (\psi_{v_\pm}(x-v_\pm t-a_\pm),
\pi_{v_\pm}(x-v_\pm t-a_\pm))
+W_0(t)\bPsi_\pm,~~~~~~t\to\pm\infty,
\ee
for solutions to
(\re{system0}) with initial data close to the solitary manifold
${\cal S}$. Here $W_0(t)$ is the dynamical group of the free wave
equation, $\bPsi_\pm$ are the corresponding {\it asymptotic
scattering states}, and the remainder converges to zero {\it in
the global energy norm}, i.e. in the norm of the Sobolev space
$\dot H^1(\R^3)\oplus L^2(\R^3)$, see Section 2. For the particle's
trajectory we prove that
\be\la{qqi} \dot q(t)\to v_\pm, ~~~~q(t)
\sim v_\pm t+a_\pm, ~~~~~~~t\to\pm\infty. \ee

The results are established under the following conditions on the charge
distribution: $\rho$ is a real-valued function of the Sobolev class $H^2(\R^3)$,
compactly supported, and spherically symmetric, i.e. \be\la{ro}
\rho,\na\rho,\na\na\rho\in L^2({\R}^3),~~~~~~~~\,\,\,\,\,\,\quad\rho (x)=0
\,\,\,\mbox{for}\,\,\,|x|\ge R_\rho, \,~~~~~~~~~~~~~\rho (x)=\rho_1(|x|). \ee We
require that all ``nonzero modes'' of the wave field are coupled to the particle.
This is formalized by the Wiener condition \be\la{W}
\hat\rho(k)=(2\pi)^{-3/2}\int\limits \, e^{ik x}\rho(x)dx\not=0 \mbox{
\,\,\,for\,\,all\,\, }k\in\R^3\setminus\{0\}\,. \ee It is the nonlinear  Fermi Golden
Rule for our model: the coupling term $\rho(x-q)$ is not orthogonal to the
eigenfunctions $e^{ikx}$ of the continuous spectrum of the linear part of the
equation (cf. \ci{BC08,BP3,Cu10,CM08,Sig93,SW3}). Note that the Wiener condition
is close to the linear version of the FGR \ci[pp. 67-68]{RS4}. As we will see, the
Wiener condition (\re{W}) is very essential for our asymptotic analysis.
 Generic examples of the coupling function $\rho$ satisfying (\re{ro}) and
 (\re{W}) are given in \ci{KSK}. In particular, the Wiener condition allows us to
identify the discrete spectral subspace corresponding to the spectral point $\lam=0$
of the linearized system. Thus, we do not impose any implicit spectral conditions
for the linearized system.
Further, we will assume that $\rho(k)$ has a fifth order zero at the point $k=0$, i.e.
\be\la{zero4}
\hat\rho^{\,\,(\al)}(0)=0\,\,\,{\rm for\,\,\,all\,\,\,multiindeces}\,\,\,\al\,\,\,
{\rm with}\,\,\,|\al|\le4.
\ee
Equivalently, the following momenta vanish:
\be\la{neutr}
\int\,x^\al\rho(x)dx=0,\,\,\,|\al|\le4,
\ee
in particular, the total charge of the particle is zero (neutrality condition).
It is easy to obtain an example of $\rho$ satisfying both (\re{W}) and (\re{zero4}).
Indeed, take a $\rho_2$ satisfying the Wiener condition (\re{W}).
Then $\rho=\De^3\rho_2$ satisfies both (\re{W}) and (\re{zero4}).

The system (\ref{system0}) describes the charged particle
interacting with
its ``own'' scalar field. The asymptotics (\ref{Si})-(\ref{qqi})
mean asymtotic stability of uniform motion,
i.e. ``the law of inertia''.
The stability is caused by ``radiative damping'', i.e. radiation of
energy to infinity appearing analytically as a local energy decay for
solutions to the linearized equation provided by
the Wiener condition (\re{ro}).
The  radiative damping was suggested first by M.Abraham in 1905
in the context of Classical Electrodynamics, \cite{A}. However,
the asymptotics (\ref{Si})-(\ref{qqi}) are not proved yet for the
Maxwell-Lorentz equations though close results are establihsed in
\ci{KS00} and \ci{IKM04}.

One could also expect asymptotics (\ref{Si})
for small perturbations of
the solitons for
the relativistic nonlinear wave equations and for the coupled
nonlinear
Maxwell-Dirac equations whose  solitons were constructed in
\ci{BL} and \ci{EGS} respectively.
Our result is a model of this situation though the relativistic
case is still open problem.

Let us briefly comment on earlier results.
In the case of weak coupling, i.e.
$\Vert\rho\Vert_{L^2}\ll1$, scattering behavior of type (\re{Si})
for the system (\re{system0})
is established
in \ci{IKSs} for all finite energy solutions.
In \ci{IKM,IKSm,IKSr} the result was extended to the cases of Klein-Gordon field,
Maxwell field, and spinning charge subject to Maxwell field respectively.
The results under the Wiener condition are not established yet.

The system (\re{system0}) under the Wiener condition was considered
in \ci{KSK,KS,KKS}. In  \ci{KSK} the convergence to stationary states
is proved for all finite energy solutions: in particular,
the relaxation of accelleration $\ddot q\to0$ as $t\to0$ holds.
In \ci{KS} the soliton-type
asymptotics for the fields is established
for all finite energy solutions.
In  \ci{KKS}
the effective dynamics under the slowly varying potential is constructed for solutions
sufficiently close to the solitary manifold.
However, the asymptotics (\re{Si}) and  (\re{qqi}) are not established in
\ci{KSK, KS, KKS}.


A long time asymptotics of type (\re{Si}) appears also in nonlinear
wave equations, like
the KDV \ci{EH81,FZ93} and the $U(1)$-invariant
nonlinear Schr\"odinger equation \ci{BP1,BP2,BP3,BS,
PW92,PW94, SW88, SW90,SW92}. In
these equations there are no particle degrees of freedom
and the solitons (\re{sosol})
correspond
to the solitary wave solutions travelling at constant velocity.

Our approach
relies on and further develops the general  strategy introduced in
the cited papers in the context
of the
$U(1)$-invariant Schr\"odinger equation.
The approach uses i) symplectic projection of the dynamics
in the Hilbert phase space onto the
symplectic orthogonal
directions to the solitary manifold to kill the  runaway
 secular solutions, ii) the modulation equations
for the motion along the  solitary manifold, and
iii) freezing of the dynamics in the nonauthonomous linearized equation.
See more details in Introduction \ci{IKV05} where
the general strategy has been developed
for the case of the Klein-Gordon equation. The
case of wave
equation
(\re{system0})
differs
significally from the  Klein-Gordon case
because of i) slow Coulombic decay of the
solitons, and ii) presence of the embedded eigenvalue in the continuous
spectrum of the linearized equation (see the comments below).

Developing the general strategy for  equations (\re{system0}),
we obtain our main result  in Sections 3-6 and 11-15 of the paper.
The main novelty in our case is thorough
establishing the appropriate
decay of the linearized dynamics
in Sections 7-10 and Appendices A, B:

I. We
do not postulate any spectral properties of the linearized equation,
calculating all the properties from the Wiener condition (\re{W}).
Namely, we show that
 i) the full zero spectral space
of the linearized equation is spanned by the
tangent vectors,
and moreover, ii)
there are no others (nonzero) discrete eigenvalues
(see Lemmas \re{Minv}, \re{sort16}
and Proposition \re{creg}).

II. Using these spectral properties,
we prove
that the linearized equation is stable in
the {\it
symplectic orthogonal complement} to the tangent space $\cT_S$
spanned by the
tangent vectors
$\pa_{a_j} S_{a,v}$ and $\pa_{v_j} S_{a,v}$, $j=1,2,3$.
We exactly
calculate in
Lemma \re{sort16}
the corresponding symplectic orthogonality conditions
for initial data of the linearized dynamics.

III. One of main peculiarities of the wave
equations
(\re{system0})
is the presence of embedded eigenvalue $\lam=0$
in the continuous
spectrum $\si_c=\R$ of the linearized equation.
This situation never happens
in all previous works on the asymptotic stability of the solitary waves
for the  Schr\"odinger and  Klein-Gordon equations.
Thus, the symplectic orthogonality condition is
imposed now at the interior point of the continuous spectrum in contrast to all
previous works in the field.
Respectively, the integrand at this point
in the spectral representation of the solution
is not smooth even if the  symplectic orthogonality condition holds.
Hence,
the integration by parts in this spectral representation
as in the case of the  Schr\"odinger and Klein-Gordon
equation,
is impossible.
For the proof of the decay in this new situation,
we transform the spectral representation
in  the proofs of Propositions \re{Q2P2} and \re{PQtdec},
and develop
 new
more subtle technique of
convolutions.

\medskip

\noindent

Our paper is organized as follows. In Section 2, we formulate the
main result. In  Section 3, we introduce the symplectic projection
onto the solitary manifold. The linearized equation is defined and
studied in Sections 4-5. In Section 6, we split the dynamics in
two components: along the solitary manifold, and in transversal
directions, and we justify the slow motion of the longitudinal
component.
Sections 7-10 concern the time decay of the transversal component
in the linearized dynamics. The time decay of the transversal component
in the nonlinear dynamics
is
established in Sections 11-14. In Section 15 we prove the main
result.
In Appendices A, B we collect routine calculations.

{\bf Acknowledgements} The authors thank V.~Buslaev for detailed
and numerous lectures on his results, H.~Spohn and E.~Kopylova for fruitful discussions.


\setcounter{equation}{0}

\section{Main Results}


\subsection{Existence of Dynamics}

 To formulate our results precisely, we need some definitions. We introduce
a suitable phase space for the Cauchy problem corresponding to
(\ref{system0}).
Let $L^2$ be the real
 Hilbert space $L^2(\R^3)$
 with the scalar product $\langle\cdot,\cdot\rangle$ and the corresponding norm $\Br\cdot\Br$, and let
 $\dot H^1$ be the completion of the real space $C_0^\infty(\R^3)$
 with the norm $\Br\nabla\psi(x)\Br$.
 Equivalently, using Sobolev embedding theorem,
 $\dot H^1=\{\psi(x)\in L^6(\R^3):~~|\nabla\psi(x)|\in L^2\}$.
Let us introduce the weighted Sobolev spaces $L^2_{\alpha}$ and $H^1_{\alpha}$ with
the norms $\Br\psi\Br_{\alpha}=\Br(1+|x|)^{\alpha}\psi\Br$ and
$\Vert\psi\Vert_{1,\al}=\Br\psi\Br_\al+\Br\na\psi\Br_\al$ respectively.

\begin{definition}\la{dE}
i)  ${\cal E}$ is the Hilbert space
 $\dot H^1\oplus L^2\oplus\R^3\oplus\R^3$ with the finite norm
 $$
  {\Vert\,Y\Vert}_{\cal E}=\Br\na\psi\Br + \Br\pi\Br + |q|  +|p|
    \quad\mbox{for}\quad Y=(\psi, \pi, q, p)\,.
   $$

 ii) ${\cal E}_{\alpha}$ is the space
 $H^1_{\alpha}\oplus L^2_{\alpha+1}\oplus {\R}^3\oplus {\R}^3$
 with the norm
 \be\la{alfa}
 \Vert Y\Vert_{\alpha}=\Vert \,Y\Vert_{{\cal E}_{\alpha}}=
 \Vert\psi\Vert_{1,\alpha} +\Br\pi \Br_{\alpha+1}+|q|+|p|.
 \ee

 iii) ${\cal F}$ is the space $\dot H^1\oplus L^2$ of fields
$F =(\psi ,\pi )$ with the finite norm
\be\la{FFF}
\Vert F \Vert_{\cal F}=\Br\na\psi\Br +\Br\pi \Br.
\ee

iv) ${\cal F}_{\alpha}$ is the space
 $H^1_{\alpha}\oplus L^2_{\alpha+1}$
 with the norm
 \be\la{Falfa}
 \Vert \,F\Vert_{\alpha}=\Vert \,F\Vert_{{\cal F}_{\alpha}}=
 \Vert\psi\Vert_{1,\alpha}+\Br\pi\Br_{\alpha+1}
 \ee
\end{definition}
 Note that
 $\dot H^1$ is not contained in $L^2$
 and for instance $\Br\psi_v\Br=\infty$ if the neutrality condition is not imposed,
see (\re{sol}) below.
 However, ${\cal E}$
 is the space of finite energy states (i.e. ${\cal H}(Y)<\infty$ for $Y \in
 \cE$) due to the following estimates which are valid for an
 arbitrary
 smooth $\psi(x)$ vanishing at infinity
 \be\la{WPHB}
 -\fr 1 {8\pi}\int\int~dx~dy~\fr{\rho(x)\rho(y)}{|x-y|}
 =
 \fr 12\langle\rho,\Delta^{-1}\rho\rangle
\leq
 \fr {1}{2}\Br\nabla\psi\Br^2+
 \langle\psi(x),\rho(x-q)\rangle
\leq
 ~~\Br\nabla\psi\Br^2-
 \fr 12\langle\rho,\Delta^{-1}\rho\rangle.
\ee
 The Hamilton functional  $\cal H$ is continuous in the space
 ${\cal E}$, and
the lower bound in (\re{WPHB}) implies that the energy (\re{hamilq0})
 is bounded from below. Note that the latter is not true if $\rho$ is
 delta-function.

 We consider the Cauchy problem for the Hamilton system (\re{system0})
 which we write as
 \be\la{WP2.1}
 \dot Y(t)=F(Y(t)),\quad t\in\R;\quad Y(0)=Y_0.
 \ee
 Here $Y(t)=(\psi(t), \pi(t), q(t), p(t))$, $Y_0=(\psi_0, \pi_0, q_0, p_0)
 $, and all
derivatives are understood in the sense of distributions.

 \begin{pro}\la{WPexistence} {\rm \ci{KSK}} Let (\re{ro}) hold.
 Then

 (i) For every $Y_0\in {\cal E}$,
the Cauchy problem (\re{WP2.1}) has a unique
 solution $Y(t)\in C(\R, {\cal E})$.

 (ii) For every $t\in\R$, the map $U(t): Y_0\mapsto Y(t)$ is continuous in
 ${\cal E}$.

 (iii)
The  energy is conserved, i.e.
 \be\la{2.4}
{\cal H}(Y(t))= {\cal H}(Y_0),\,\,\,\,\,t\in\R,
 \ee
 and the velocity is bounded:
 \be\la{WP2.1'}
 |\dot q(t)|\leq \ov v<1,~~~~t\in\R,
 \ee
 with
 some $\ov v$ which depends on $Y_0$.
 \end{pro}


\subsection{Solitary Manifold and Main Result}

Let us compute the  solitons (\re{sosol}).
The substitution to (\re{system0}) gives the following stationary equations,
\be\la{stfch}
\left.\ba{rclrcl}
-v\cdot\na\psi_v(y)&=&\pi_v(y)~,&-v \cdot\na\pi_v(y)&=&\De\psi_v(y)
-\rho(y)
\\\\
v&=&\ds\fr{p_v}{\sqrt{1+p_v^2}}~,&0&=&-\ds\int\na\psi_v(y)\rho(y)\,dy
\ea\right|
\ee
Then the first two equations imply
\be\la{Lpv}
\Lam_v\psi_v(y):=(-\De+(v\cdot\na)^2)\psi_v(y)=-\rho(y),~~~~~~~~y\in\R^3.
\ee
For $|v|<1$ the equation (\re{Lpv})
defines a unique function $\psi_v\in\dot H^1(\R^3)$.
If $v$ is given and $|v|<1$, then $p_v$ can be found from the third equation of (\re{stfch}).
 Further, functions $\rho$ and $\psi_v$ are even due to (\re{ro}).
Thus, $\na\psi_v$ is odd and the last equation of (\re{stfch})
holds. Hence, the soliton solution (\re{sosol}) exists and is
defined uniquely for any couple $(a,v)$ with $|v|<1$.

The soliton can be computed by the Fourier transform
$\hat\psi(k):=(2\pi)^{-3/2}\ds\int e^{ikx}\psi(x)dx$:
\be\la{solF}
\hat\psi_v(k)=-\fr{\hat\rho(k)}{k^2-(kv)^2},~~~~~~
\hat\pi_v(k)=-\fr{ikv\hat\rho(k)}{k^2-(kv)^2}
\ee
In the coordinate space
\be\la{sol} \psi_v(x)=
 -\fr  1{4\pi} \int
\fr {\rho(y)d^3y} {|\ga(y-x)_\|+(y-x)_\bot|}\,,
\,\,\,
\pi_v(x) =-v\cdot\na\psi_v(x) , \,\,\, p_v=\ga v. \ee Here we
set $\ga=1/\sqrt{1-v^2}$ and $x=x_\Vert+x_\bot$, where
$x_\Vert\Vert v$ and $x_\bot\bot v$ for $x\in\R^3$.
From the condition (\re{neutr}) it follows that
$$
\psi_v(y)\sim|y|^{-6},\,\,\,\pi_v(y)\sim|y|^{-7}\,\,\,{\rm as}\,\,\,|y|\to\infty
$$
and thus,
\be\la{psipidec}
\psi_v\in H^1_{\alpha},\,\,\,\alpha<9/2;\,\,\,\pi_v\in L^2_{\alpha},\,\,\,\alpha<11/2.
\ee

\begin{definition}
A soliton state is $S(\si):=(\psi_v(x-b),\pi_v(x-b),b,p_v)$, where
 $\si:=(b,v)$ with
 $b\in\R^3$ and $|v|<1$.
\end{definition}
By (\re{psipidec}) for the soliton states we have
\be\la{solal}
S(\si)\in{\cal E}_\alpha,\,\,\,\alpha<\fr92.
\ee
Obviously, the soliton solution (\re{sosol})
admits the representation $Y_{a,v}(t)=S(\si(t))$, where
\be\la{sigma}
\si(t)=(b(t),v(t))=(vt+a,v).
\ee
\begin{definition}
The solitary manifold is the set ${\cal S}:=\{S(\si):b\in \R^3,
|v|<1\}$.
\end{definition}

The main result of our paper is the following theorem.
\begin{theorem}\la{main}
Let (\re{ro}), the Wiener condition (\re{W}), and the condition (\re{zero4}) hold.
Let $0<\de<1/2$, set $\beta:=4+\de$. Consider the solution $Y(t)$ to the Cauchy problem
(\re{WP2.1}) with the initial state $Y_0$ which is sufficiently
close to the solitary manifold: \be\la{close}
Y_0=S(\si_0)+Z_0,\,\,\,\,d_\beta:=\Vert Z_0\Vert_{\beta}\ll 1. \ee
Then the asymptotics hold for $t\to\pm\infty$, \be\la{qq} \dot
q(t)=v_\pm+{\cal O}(|t|^{-1-\de}), ~~~~q(t)=v_\pm t+a_\pm+{\cal
O}(|t|^{-2\de}), \ee \be\la{S} (\psi(x,t), \pi(x,t))=
(\psi_{v_\pm}(x-v_\pm t-a_\pm), \pi_{v_\pm}(x-v_\pm t-a_\pm))
+W_0(t)\bPsi_\pm+r_\pm(x,t) \ee with \be\la{rm} \Vert
r_\pm(t)\Vert_\cF=\cO(|t|^{-\de}). \ee
\end{theorem}

\smallskip

It suffices to  prove the asymptotics (\re{qq}), (\re{S}) for $t\to+\infty$
since the system (\re{system0}) is time reversible.


\setcounter{equation}{0}

\section{Symplectic Projection}

\subsection{Symplectic Structure and Hamilton Form}

The system (\re{system0}) reads as the Hamilton system
\be\la{ham}
\dot Y=J{\cal D}{\cal H}(Y),\,\,\,J:=\left(
\ba{cccc}
0 & 1 & 0 & 0\\
-1 & 0 & 0 & 0\\
0 & 0 & 0 & I_3\\
0 & 0 & -I_3 & 0\\
\ea
\right),\,\,Y=(\psi,\pi,q,p)\in{\cal E},
\ee
where ${\cal D}{\cal H}$ is the Fr\'echet derivative of
 the Hamilton functional (\re{hamilq0}), $I_3$ is the $3\times3$ identity matrix.
Let us identify the tangent space to ${\cal E}$, at every point,
 with ${\cal E}$.
Consider the bilinear form $\Om$
 defined on
${\cal E}$ by $\Om=\ds\int d\psi(x)\we d\pi(x)\,dx+dq\we dp$, i.e.
\be\la{OmJ}
\Om(Y_1,Y_2)=\langle Y_1,JY_2\rangle,\,\,\,Y_1,Y_2\in {\cal E},
\ee
where
$$
\langle Y_1,Y_2\rangle:=\langle\psi_1,\psi_2\rangle+
\langle\pi_1,\pi_2\rangle+q_1  q_2+p_1  p_2
$$
and $\langle\psi_1,\psi_2\rangle$ stands for the scalar product
$\ds\int\psi_1(x)\psi_2(x)dx$ or its different extensions.
\begin{definition}
i) $Y_1\nmid Y_2$ means that $Y_1$ is symplectic orthogonal to $Y_2$, i.e. $\Om(Y_1,Y_2)=0$.

ii) A projection operator $\bP:{\cal E}\to{\cal E}$
is called symplectic
 orthogonal if $Y_1\nmid Y_2$ for $Y_1\in\mbox{\rm Ker}\5\bP$ and
$Y_2\in\mbox \Im\bP$.
\end{definition}


\subsection{Symplectic Projection onto Solitary Manifold}

From now on we suppose that the condition (\re{zero4}) is satisfied.
Let us consider the tangent space $\cT_{S(\si)}{\cal S}$ to
the manifold ${\cal S}$ at a point $S(\si)$, where $\si=(b,v)$.
The vectors $\tau_j:=\pa_{\si_j}S(\si)$, where $\pa_{\si_j}:=
\pa_{b_j}$ and $\pa_{\si_{3+j}}:=\pa_{v_{j}}$ with $j=1,2,3$,
form a basis in $\cT_{\si}{\cal S}$. In detail,
\be\la{inb}
\left.\ba{rclrrrrcrcl}
\tau_j=\tau_j(v)&:=&\pa_{b_j}S(\si)=
(&\!\!\!\!-\pa_j\psi_v(y)&\!\!\!\!,&\!\!\!\!-\pa_j\pi_v(y)&\!\!\!\!,
&\!\!e_j&\!\!\!\!,&\!\!0&\!\!\!\!)
\\
\tau_{3+j}=\tau_{3+j}(v)&:=&\pa_{v_j}S(\si)=(&\!\!\!\!\pa_{v_j}\psi_v(y)
&\!\!\!\!,&\!\!\!\!
\pa_{v_j}\pi_v(y)&\!\!\!\!,&\!\!0&\!\!\!\!,&\!\!
\pa_{v_j}p_v&\!\!\!\!) \ea\right|~~~j=1,2,3, \ee where  $y:=x-b$
is the ``coordinate in the moving frame'', $e_1=(1,0,0)$ etc.

By (\re{psipidec}) for the tangent vectors we have
\be\la{tangal}
\tau_j(v)\in\cE_\al,\,\,\alpha<\fr92;\,\,\,j=1,\dots,6.
\ee

\begin{lemma}\la{Ome}
The matrix with the elements $\Om(\tau_l(v),\tau_j(v))$ is
non-degenerate for $|v|<1$.
\end{lemma}
The proof is made by a straightforward computation, see \ci{IKV05}, Lemma 3.2
for the case $m=0$ (note that the left hand side of the identity \ci[(A.8)]{IKV05}
is well defined by the condition (\re{zero4})).

Now we show that in a small neighborhood of the soliton
manifold ${\cal S}$ a ``symplectic orthogonal projection''
onto ${\cal S}$ is well defined. Let us introduce the translations
 $T_a:(\psi(x),\pi(x),q,p)\mapsto
(\psi(x-a),\pi(x-a),q+a,p)$, $a\in\R^3$. Note that the
manifold ${\cal S}$ is invariant with respect to the translations.
\begin{definition}Let us denote by $v(Y):=p/\sqrt{1+p^2}$ where
$p\in\R^3$ is the last component of the vector $Y$.
\end{definition}

\begin{lemma}\la{skewpro}
Let (\re{ro}) hold, $\al>-9/2$ and $\ov v<1$.
Then
\\
i) there exists a neighborhood ${\cal O}_\al({\cal S})$ of ${\cal
S}$ in ${\cal E}_\al$ and a map $\bPi:{\cal O}_\al({\cal
S})\to{\cal S}$  such that $\bPi$ is uniformly continuous on
${\cal O}_\al({\cal S})\cap\{Y\in\cE_\al:v(Y)\le\ov v\}$ in the
metric of ${\cal E}_\al$, \be\la{proj} \bPi Y=Y~~\mbox{for}~~
Y\in{\cal S}, ~~~~~\mbox{and}~~~~~ Y-S \nmid \cT_S{\cal
S},~~\mbox{where}~~S=\bPi Y. \ee ii) ${\cal O}_\al({\cal S})$ is
invariant with respect to the translations
 $T_a$, and
\be\la{commut} \bPi T_aY=T_a\bPi Y,~~~~~\mbox{for}~~Y\in{\cal
O}_\al({\cal S}) ~~\mbox{and}~~a\in\R^3. \ee iii) For any $\ov
v<1$ there exists a $\ti v<1$ such that
 $|v(\bPi Y)|<\ti v$ when  $|v(Y)|<\ov v$.
\\\\
iv) For any $\ti v<1$ there exists an $r_\al(\ti v)>0$ s.t.
$S(\si)+Z\in\cO_\al(\cS)$ if $|v(S(\si))|<\ti v$ and $\Vert
Z\Vert_\al<r_\al(\ti v)$.
\end{lemma}
The proof is similar to that of \ci[Lemma 3.4]{IKV05}.

\medskip

We will call $\bPi$ the {\it symplectic orthogonal projection} onto ${\cal S}$.

\bc {\rm \ci[Corollary 3.5]{IKV05}}.
The condition (\re{close})
implies that $Y_0=\ti S+\ti Z_0$ where $\ti S=\bPi Y_0$, and
\be\la{closeZ}
\Vert\ti Z_0\Vert_{\beta}\ll 1.
\ee
\ec


\setcounter{equation}{0}

\section{Linearization on the Solitary Manifold}

Let us consider a solution to the system (\re{system0}), and split it as
 the sum

\be\la{dec} Y(t)=S(\si(t))+Z(t), \ee where
$\si(t)=(b(t),v(t))\in\R^3\times\{|v|<1\}$ is an arbitrary smooth
function of
 $t\in\R$.
In detail, denote $Y=(\psi,\pi,q,p)$ and $Z=(\Psi,\Pi,Q,P)$.
Then (\re{dec}) means that
\be \la{add}
\left.
\ba{rclrcl}
\psi(x,t)&=&\psi_{v(t)}(x-b(t))+\Psi(x-b(t),t),
&q(t)&=&b(t)+Q(t)\\
\pi(x,t)&=&\5\pi_{v(t)}(x-b(t))+\Pi(x-b(t),t),
&p(t)&=&p_{v(t)}+P(t)
\ea
\right|
\ee
Let us
substitute (\re{add}) to (\re{system0}), and
linearize the equations in $Z$. Later we
 will choose $S(\si(t))=\bPi Y(t)$, i.e. $Z(t)$ is symplectic
orthogonal to $\cT_{S(\si(t))}{\cal S}$. However, this orthogonality
condition is not needed for the formal process of linearization.
The orthogonality condition will be important in Section 6, where we derive
``modulation equations'' for the parameters $\si(t)$.

Let us proceed to linearization. Setting $y=x-b(t)$ which is
the {\it coordinate in the moving frame}, we obtain from
(\re{add}) and (\re{system0}) that
\be\la{addeq}
\left.
\ba{rcl}
\dot\psi&=&\dot v\cdot \na_v\psi_{v(t)}(y)-\dot b\cdot \na\psi_{v(t)}(y)+
\dot\Psi(y,t)-\dot b\cdot \na\Psi(y,t)=\pi_{v(t)}(y)+\Pi(y,t)
\\\\
\dot\pi&=&\dot v\cdot \na_v\pi_{v(t)}(y)-\dot b \cdot\na\pi_{v(t)}(y)+
\dot\Pi(y,t)-\dot b\cdot \na\Pi(y,t)
\\\\
&=&\De\psi_{v(t)}(y)+\De\Psi(y,t)-\rho(y-Q)
\\\\
\dot q&=&\dot b+\dot Q=\ds\fr{p_v+P}{\sqrt{1+(p_v+P)^2}}
\\\\
\dot p&=&\dot v\cdot \na_v p_{v(t)}+\dot
P=-\langle\na(\psi_{v(t)}(y)+ \Psi(y,t)),\rho(y-Q)\rangle \ea
\right| \ee The equations are linear in $\Psi$ and $\Pi$, hence it
remains to extract linear terms in $Q$ and $P$.
Within this section, we estimate the remainders in the norms of the space ${\cal E}_{\al}$
with an arbitrary $\al>0$.

First note that
$\rho(y-Q)=\rho(y)-Q \cdot\na\rho(y)-N_2(Q)$, where
$-N_2(Q)=\rho(y-Q)-\rho(y)+Q\cdot \na\rho(y)$; for $N_2(Q)$ the
bound holds, \be\la{N2} \Br N_2(Q)\Br_{\al}\le C(\ov
Q)Q^2 \ee uniformly in $|Q|\le\ov Q$ for any fixed $\ov Q$. Second, the
Taylor expansion gives
$$
\fr{p_v+P}{\sqrt{1+(p_v+P)^2}}=v+\nu(P-v(v \cdot P))+N_3(v,P),
$$
where $\nu=\nu_v:=(1+p_v^2)^{-1/2}=\sqrt{1-v^2}$, and
\be\la{N3}
|N_3(v,P)|\le C(\ti v)P^2
\ee
 uniformly in $v$ with $|v|\le\ti v<1$.
Using the equations (\re{stfch}),  we obtain from (\re{addeq})
the following equations for the components of the vector $Z(t)$:
\be\la{Phi}
\left.\ba{lcl}
\!\!\!\!\dot \Psi(y,t)\!\!\!\!&=&\!\!\!\!\Pi(y,t)+\dot b\cdot \na\Psi(y,t)+
(\dot b-v)\cdot \na\psi_v(y)-\dot v \cdot\na_v\psi_v(y)
\\\\
\!\!\!\!\dot \Pi(y,t)\!\!\!\!&=&\!\!\!\!\De\Psi(y,t)\!+\!
\dot b\cdot \!\na\Pi(y,t)\!+\!
Q \!\cdot\!\na\rho(y)\!+\!(\dot b\!-\!v)\!\cdot\! \na\pi_v(y)\!-
\!\dot v\cdot \!\na_v\pi_v(y)\!+\!N_2
\\\\
\!\!\!\!\dot Q(t)\!\!\!\!&=&\!\!\!\!\nu_v(I_3-v\otimes v)P+(v-\dot b)+N_3
\\\\
\!\!\!\!\dot
P(t)\!\!\!\!&=&\!\!\!\!\langle\Psi(y,t),\na\rho(y)\rangle+
\langle\na\psi_v(y), Q \cdot\na\rho(y)\rangle-\dot v\cdot \na_v
p_v+N_4(v,Z) \ea\right| \ee where
$N_4(v,Z)=\langle\na\psi_v,N_2(Q)\rangle+ \langle\na\Psi,Q\cdot
\na\rho\rangle+\langle\na\Psi,N_2(Q)\rangle$. Clearly, $N_4(v,Z)$
satisfies the following estimate \be\la{N4} |N_4(v,Z)|\le
C_\beta(\rho,\ti v,\ov Q)\Big[Q^2+ \Vert\Psi\Vert_{-\al}|Q|
 \Big],
\ee
uniformly in $v,Q$ with $|v|\le \ti v$ and $|Q|\le \ov Q$.
We can write the equations (\re{Phi}) as
\be\la{lin}
\dot Z(t)=A(t)Z(t)+T(t)+N(t),\,\,\,t\in\R.
\ee
Here the operator $A(t)$ depends on $\si(t)=(b(t),v(t))$. We will
use the parameters $v=v(t)$ and $w:=\dot b(t)$. Then $A(t)$ can be
written in the form \be\la{AA} A(t)\left( \ba{c} \Psi \\ \Pi \\ Q
\\ P \ea \right)=A_{v,w}\left( \ba{c} \Psi \\ \Pi \\ Q \\ P \ea
\right):=\left( \ba{cccc}
w \cdot\na & 1 & 0 & 0 \\
\De & w\cdot \na & \na\rho\cdot & 0 \\
0 & 0 & 0 & B_v \\
\langle\cdot,\na\rho\rangle & 0 & \langle\na\psi_v,\cdot\na\rho\rangle & 0
\ea
\right)\left(
\ba{c}
\Psi \\ \Pi \\ Q \\ P
\ea
\right),
\ee
where $B_{v}=\nu_{v}(I_3-v\otimes v)$. Furthermore,   $T(t)$ and $N(t)$ in (\re{lin})
stand for \be\la{TN} T(t)=T_{v,w}=\left( \ba{c}
(w-v)\cdot\na\psi_v-\dot v\cdot\na_v\psi_v\\
(w-v)\cdot\na\pi_v-\dot v\cdot\na_v\pi_v\\
v-w \\
-\dot v\cdot\na_v p_v \ea \right),\,\,\,\,N(t)=N(\si,Z)=\left(
\ba{c} 0 \\ N_2(Z) \\ N_3(v,Z) \\ N_4 (v,Z) \ea \right), \ee where
$v=v(t)$, $w=w(t)$, $\si=\si(t)=(b(t),v(t))$, and $Z=Z(t)$. Since $|Q|\le\Vert Z\Vert_{-\al}$
for any $\al$, the
estimates (\re{N2}) with $\ov Q=r_{-\al}(\ti v)$, (\re{N3}) and (\re{N4}) imply the following
\bl\la{Nbeta}
For any $\al>0$
\be\la{N14} \Vert N(\si,Z)\Vert_\al\le C(\ti v) \Vert
Z\Vert_{-\al}^2 \ee uniformly in $\si,Z$ with $\Vert
Z\Vert_{-\al}\le r_{-\al}(\ti v)$ and
$|v|<\ti v$.
\el
\brs\la{rT} {\rm i) The term $A(t)Z(t)$ in the right
hand side of the equation  (\re{lin}) is linear  in $Z(t)$, and
$N(t)$ is a {\it high order term} in $Z(t)$. On the other hand,
$T(t)$ is a {\it zero order term} which does not vanish at $Z(t)=0$
since $S(\si(t))$ generally is not a soliton solution if
(\re{sigma}) does not hold (though $S(\si(t))$ belongs to the
solitary manifold).
\\
ii) Formulas (\re{inb}) and (\re{TN}) imply:
\be\la{Ttang}
T(t)=-\sum\limits_{l=1}^3[(w-v)_l\tau_l+\dot v_l\tau_{l+3}]
\ee
and hence $T(t)\in \cT_{S(\si(t))}{\cal S}$, $t\in\R$.
This fact suggests an unstable character of the nonlinear dynamics
{\it along the solitary manifold}.
}
\ers


\setcounter{equation}{0}
\section{The Linearized Equation}

Here we collect some Hamiltonian and spectral properties of the
operator (\re{AA}). The statements of the section
are particular cases of those in \ci{IKV05}, Section 5 for $m=0$. First, we
consider the linear equation \be\la{line} \dot
X(t)=A_{v,w}X(t),~~~~~~~t\in\R \ee with an arbitrary fixed $v$
such that $|v|<1$, and $w\in \R^3$. Let us define the space
$$
{\cal E}^+=H^2(\R^3)\oplus H^1(\R^3)\oplus\R^3\oplus\R^3.
$$

\begin{lemma} \la{haml}
i) For any $v$, $|v|<1$ and $w\in \R^3$ the equation (\re{line})
formally
 can be written as the Hamilton system  (cf. (\re{ham})),
\be\la{lineh}
\dot X(t)=
JD{\cal H}_{v,w}(X(t)),~~~~~~~t\in\R,
\ee
where $D{\cal H}_{v,w}$ is the Fr\'echet derivative of the
Hamilton functional
\beqn
{\cal H}_{v,w}(X)&=&\fr12\int\Big[|\Pi|^2+
|\na\Psi|^2\Big]dy+\int\Pi w\cdot\na\Psi dy+\int\rho(y)
 Q\cdot\na\Psi dy
\nonumber\\
&+&
\fr12P\cdot B_vP-\fr12\langle Q\cdot\na\psi_{v}(y),Q\cdot\na\rho(y)\rangle,
~~~~X=(\Psi,\Pi,Q,P)\in \cE.
\la{H0}
\eeqn
\\
ii) Energy conservation law holds
for the solutions $X(t)\in C(\R,\cE)$,
\be\la{enec}
\cH_{v,w}(X(t))=\co,~~~~~t\in\R.
\ee
iii) The skew-symmetry relation holds,
\be\la{com}
\Omega(A_{v,w}X_1,X_2)=-\Omega(X_1,A_{v,w}X_2), ~~~~X_1\in \cE,\,\,\,X_2\in \cE^+.
\ee
\end{lemma}

\begin{lemma} \la{ljf}
The operator $A_{v,w}$ acts on the
tangent vectors $\tau_j(v)$ to the solitary manifold
as follows,
\be\la{Atanform}
A_{v,w}[\tau_j(v)]=(w-v)\cdot\na\tau_j(v),\,\,\,A_{v,w}[\tau_{j+3}(v)]=
(w-v)\cdot\na\tau_{j+3}(v)+\tau_j(v),\,\,\,j=1,2,3.
\ee
\end{lemma}

We will apply Lemmas \re{haml} and \re{ljf} mainly to the operator $A_{v,v}$
corresponding to $w=v$.
In that case (\re{Atanform}) reads
\be\la{Atanformv}
A_{v,v}[\tau_j(v)]=0,\,\,\,A_{v,v}[\tau_{j+3}(v)]=
\tau_j(v),\,\,\,j=1,2,3.
\ee
Moreover, the linearized equation acquires the
additional essential feature.

\bl\la{ceig} Let us assume that $w=v$ and $|v|<1$. Then the Hamilton functional (\re{H0})
reads, see {\rm\ci[(5.14)]{IKV05}}
\be\la{H0vv}
{\cal H}_{v,v}(X)=\fr12\int\left(|\Pi+v\cdot\na\Psi|^2+|\Lam_v^{1/2}
\Psi-\Lam_v^{-1/2}Q\cdot\na\rho|^2\right)dx+\fr12P\cdot B_vP\ge 0.
\ee
Here $\Lam_v$ is the operator defined by (\re{Lpv}).
\el

\br {\rm Lemma \re{ceig}  together with energy conservation
(\re{enec}) imply the analyticity of the resolvent
$(A_{v,v}-\lam)^{-1}$ for $\Re\lam>0$, see below.}\er

\br
{\rm
 For a soliton solution of the system
(\re{system0}) we have $\dot b=v$, $\dot v=0$, and hence $T(t)\equiv 0$.
Thus, the equation
(\re{line})
is the linearization of the system
(\re{system0}) on a soliton solution. In fact, we do not linearize
 (\re{system0}) on a soliton solution, but on a trajectory
$S(\si(t))$ with $\si(t)$
 being nonlinear in $t$. We will show later that $T(t)$ is quadratic in $Z(t)$
if we choose
$S(\si(t))$ to be  the symplectic orthogonal projection of $Y(t)$.
Then
(\re{line})  is again the linearization of (\re{system0}).
}
\er


\setcounter{equation}{0}

\section{Symplectic Decomposition of the Dynamics}

Here we decompose the dynamics in two components: along the manifold
 ${\cal S}$ and in transversal directions. The equation (\re{lin})
is obtained without any assumption on $\si(t)$ in (\re{dec}). We are going to choose
$S(\si(t)):=\bPi Y(t)$, but then we need to know  that \be\la{YtO} Y(t)\in
\cO_\al(\cS),~~~~~t\in\R, \ee with some $\cO_\al(\cS)$ defined in Lemma \re{skewpro}.
It is true for $t=0$ and $\al\le\beta:=4+\de$ by (\re{closeZ}),
if $d_\beta>0$ in
(\re{close}) is sufficiently small.
Then  $S(\si(0))=\bPi Y(0)$ and $Z(0)=Y(0)-S(\si(0))$ are well defined.

We set $\al=-\beta$ and will prove below that (\re{YtO}) holds
if $d_\beta$ is sufficiently small.
First, the a priori estimate (\re{WP2.1'})
together with Lemma \re{skewpro} iii)
imply that $\bPi Y(t)=S(\si(t))$ with $\si(t)=(b(t),v(t))$,
 and
\be\la{vsigmat}
|v(t)|\le \ti v<1,~~~~~~t\in\R
\ee
if $Y(t)\in \cO_{\al}(\cS)$. Denote by $r_{\al}(\ti v)$ the positive
number from
Lemma \re{skewpro} iv) which corresponds to the chosen $\al=-\beta$.
Then $S(\si)+Z\in \cO_{\al}(\cS)$ if
$\si=(b,v)$ with $|v|<\ti v$ and
$ \Vert Z\Vert_{\al}<r_{\al}(\ti v)$.
Note that (\re{closeZ}) implies
$\Vert Z(0)\Vert_{\al}<r_{\al}(\ti v)$ if
 $d_\beta$ is sufficiently small.
Therefore, $S(\si(t))=\bPi Y(t)$ and  $Z(t)=Y(t)-S(\si(t))$
are well defined for
$t\ge 0$
so small that
 $\Vert Z(t)\Vert_{\al} < r_{\al}(\ti v)$.
This is formalized by the following standard definition.
\begin{definition}
$t_*$ is the ``exit time'', \be\la{t*} t_*=\sup \{t>0: \Vert
Z(s)\Vert_{-\beta} < r_{-\beta}(\ti v),~~0\le s\le
t\},\,\,\,Z(s)=Y(s)-S(\si(s)). \ee
\end{definition}
One of our main goals is to prove that $t_*=\infty$
if $d_\beta$ is sufficiently small.
This would follow if we show that
\be\la{Zt}
\Vert Z(t)\Vert_{-\beta}<r_{-\beta}(\ti v)/2,~~~~~0\le t < t_*.
\ee
Now $N(t)$ in (\re{lin})
satisfies, by (\re{N14}) with $\al=\beta$, the following estimate:
\be\la{Nest}
\Vert N(t)\Vert_{\beta}\le C(\ti v)\Vert Z(t)\Vert^2_{-\beta},
\,\,\,0\le t<t_*.
\ee


\subsection{Longitudinal Dynamics: Modulation Equations}

From now on
 we fix the decomposition
$Y(t)=S(\si(t))+Z(t)$ for $0<t<t_*$
by setting $S(\si(t))=\bPi Y(t)$ which is equivalent to the
 symplectic orthogonality condition of type (\re{proj}),
\be\la{ortZ}
Z(t)\nmid\cT_{S(\si(t))}{\cal S},\,\,\,0\le t<t_*.
\ee
This allows us to simplify drastically the
asymptotic analysis of the
dynamical equations (\re{lin})
 for the transversal component $Z(t)$. As the first step, we derive the
longitudinal dynamics, i.e. the {\it modulation equations} for the parameters
 $\si(t)$.
Let us derive a system of ordinary differential equations for the
vector $\si(t)$. For this purpose,
let us write (\re{ortZ}) in the form
\be\la{orth}
\Om(Z(t),\tau_j(t))=0,\,\,j=1,\dots,6, ~~~~~~~0\le t<t_*,
\ee
where the vectors $\tau_j(t)=\tau_j(\si(t))$ span the
tangent space
$\cT_{S(\si(t))}{\cal S}$.
Note that $\si(t)=(b(t),v(t))$, where
\be\la{sit}
|v(t)|\le \ti v<1,~~~~~~~~~0\le t<t_*,
\ee
by Lemma \re{skewpro} iii).
  It would be convenient for us to use some other
parameters $(c,v)$ instead of $\si=(b,v)$, where
$c(t)=
b(t)-\ds\int^t_0 v(\tau)d\tau$ and
\be\la{vw}
\dot c(t)=\dot b(t)-v(t)=w(t)-v(t), ~~~~~~~~~0\le t<t_*.
\ee
We do not
need an explicit form of the equations for $(c,v)$ but the following
 statement.

\begin{lemma}\la{mod}
Let $Y(t)$ be a solution to the Cauchy problem (\re{WP2.1}), and (\re{zero4}),
(\re{dec}), (\re{orth}) hold. Then $(c(t),v(t))$ satisfies the
equation \be\la{parameq} \left( \ba{l} \dot c(t) \\ \dot v(t) \ea
\right)={\cal N}(\si(t),Z(t)), ~~~~~~~0\le t<t_*, \ee where
\be\la{NZ} {\cal N}(\si,Z)={\cal O}(\Vert Z\Vert^2_{-\beta}) \ee
uniformly in  $\si\in\{(b,v):|v|\le\ti v\}$.
\end{lemma}
\pru We differentiate (\re{orth}) in $t$ and take the equation (\re{lin})
into account. Then (see details of computation in \ci{IKV05}, Lemma 6.2) we obtain,
in the vector form \ci[(6.18)]{IKV05}:
\be\la{modular}
0=\Om(v)\left(
\ba{c}
\dot c \\ \dot v
\ea
\right)+{\cal M}_0(\si,Z)\left(
\ba{c}
\dot c \\ \dot v
\ea
\right)+{\cal N}_0(\si,Z),\,\,\,{\cal N}_{0j}(\si,Z)=\Om(N,\tau_j).
\ee
Here the matrix $\Om(v)$ has the matrix elements $\Om(\tau_l,\tau_j)$ and hence
is invertible by Lemma \re{Ome}. The $6\times6$ matrix ${\cal M}_0(\si,Z)$ has
the matrix elements $\sim\,\,\Vert Z\Vert_{-\beta}$ and hence we can resolve the
equation (\re{modular}) with respect to $(\dot c,\dot v)$. Then (\re{NZ})
follows from Lemma \re{Nbeta} with $\al=\beta$, since ${\cal N}_0
={\cal O}(\Vert Z\Vert_{-\beta}^2)$. \hfill $\Box$

\subsection{Decay for the Transversal Dynamics}

In Section 15 we will show that our main Theorem \re{main}
can be derived from the following time decay of the
transversal component $Z(t)$:
\bp\la{pdec}
 Let all conditions of Theorem \re{main} hold. Then $t_*=\infty$, and
\be\la{Zdec} \Vert Z(t)\Vert_{-\beta}\le \ds\fr {C(\rho,\ov
v,d_\beta)}{(1+t)^{1+\de}},~~~~~t\ge0. \ee \ep We will derive
(\re{Zdec}) in Sections 8 -- 14 from our equation (\re{lin}) for the
transversal component $Z(t)$. This equation can be specified using
Lemma \re{mod}. Namely, by (\re{Ttang}) and (\re{vw})
$$
T(t)=-\sum\limits_{l=1}^3[\dot c_l\tau_l+\dot v_l\tau_{l+3}].
$$
Note that the norm $\Vert T(t)\Vert_{\beta}$ is well-defined by (\re{tangal}).
Then Lemma \re{mod} implies
\be\la{Tta}
\Vert T(t)\Vert_{\beta}\le C(\ti v)\Vert Z(t)\Vert^2_{-\beta},
~~~~~~~~~0\le t<t_*.
\ee
Thus, in
(\re{lin}) we should combine the terms $T(t)$ and $N(t)$ and obtain
\be\la{reduced} \dot
Z(t)=A(t)Z(t)+\ti N(t), ~~~~~~~~~0\le t<t_*, \ee where
$A(t)=A_{v(t),w(t)}$, and $\ti N(t):=T(t)+N(t)$. From (\re{Tta}) and (\re{Nest}) we obtain that
\be\la{redN} \Vert\ti  N(t)\Vert_{\beta}\le C(\ti v)\Vert
Z(t)\Vert^2_{-\beta},~~~~ ~~~~~~~~~0\le t<t_*. \ee

In the
remaining part of our paper we will analyze mainly the {\bf basic
equation} (\re{reduced}) to establish the decay (\re{Zdec}). We
are going to derive the decay using the bound (\re{redN}) and
 the
orthogonality condition  (\re{ortZ}).

Let us comment on two main difficulties in proving
(\re{Zdec}). The difficulties are common for the problems
studied in \ci{BP2,Cu}.
First, the linear part of the equation is non-autonomous,
hence we cannot apply directly known methods of scattering theory.
Similarly to the approach of  \ci{BP2,Cu},
we reduce the problem to
 the analysis of the
 {\it frozen} linear equation,
\be\la{Avv} \dot X(t)=A_1X(t), ~~t\in\R, \ee where $A_1$ is the operator
$A_{v_1,v_1}$ defined in (\re{AA}) with $v_1=v(t_1)$ and a fixed $t_1\in[0,t_*)$.
Then we estimate the error by the method of majorants. Let us note that recently some
methods of freezing were developed by Cuccagna and Mizumachi, \ci{Cu10,CM08}.

Second, even for the frozen equation (\re{Avv}),
the decay
of type  (\re{Zdec}) for all solutions does not hold
 without  the
orthogonality condition  of type (\re{ortZ}). Namely, by
(\re{Atanformv}) the equation (\re{Avv}) admits the {\it secular
solutions} \be\la{secs} X(t)=\sum_1^3 C_{j}\tau_j(v_1)+\sum_1^3
D_j[\tau_j(v_1)t+\tau_{j+3}(v_1)] \ee which arise also by
differentiation of the soliton (\re{sosol}) in the parameters $a$
and $v_1$ in the moving coordinate $y=x-v_1t$. Hence, we have to
take into account the orthogonality condition  (\re{ortZ}) in
order to avoid the secular solutions. For this purpose we will
apply the corresponding symplectic orthogonal projection which
kills the ``runaway solutions''  (\re{secs}).

\br
{\rm
The solution (\re{secs}) lies in the tangent space  $\cT_{S(\si_1)}{\cal S}$
with
$\si_1=(b_1,v_1)$ (for an arbitrary $b_1\in\R$)
that
suggests an unstable character of the nonlinear dynamics
{\it along the solitary manifold} (cf. Remark \re{rT} ii)).
}
\er

\bd i) Denote by $\bPi_v$, $|v|<1$, the symplectic orthogonal
projection of ${\cal E}$ onto the tangent space $\cT_{S(\si)}{\cal
S}$, and
 $\bP_v=\bI-\bPi_v$.
\\
ii) Denote by $\cZ_v=\bP_v\cE$ the space symplectic
orthogonal to $\cT_{S(\si)}{\cal S}$ with
$\si=(b,v)$ (for an arbitrary $b\in\R$).
\ed
Note that by the linearity,
\be\la{Piv}
\bPi_vZ=\sum\bPi_{jl}(v)
\tau_j(v)\Om(\tau_l(v),Z),~~~~~~~~~~Z\in\cE,
\ee
 with some smooth coefficients $\bPi_{jl}(v)$.
Hence, the projector $\bPi_v$, in the variable $y=x-b$,
does not depend on $b$, and this explains the choice
of the subindex in $\bPi_v$ and $\bP_v$.
Now we have the symplectic orthogonal decomposition
\be\la{sod}
\cE=\cT_{S(\si)}{\cal S}+\cZ_v,~~~~~~~\si=(b,v),
\ee
and the symplectic orthogonality  (\re{ortZ})
can be written in the following equivalent forms,
\be\la{PZ}
\bPi_{v(t)} Z(t)=0,~~~~\bP_{v(t)}Z(t)= Z(t),~~~~~~~~~0\le t<t_*.
\ee

\br\la{rZ}
{\rm
The tangent space $\cT_{S(\si)}{\cal S}$ is invariant under
the operator $A_{v,v}$ by Lemma \re{ljf}, hence
the space  $\cZ_v$ is also invariant by
(\re{com}): $A_{v,v}Z\in \cZ_v$
for {\it sufficiently smooth}  $Z\in \cZ_v$.
}
\er

The following
 proposition is one of main ingredients for proving
(\re{Zdec}). Let us consider the Cauchy problem for the  equation
(\re{Avv}) with $A=A_{v,v}$ for a fixed $v$, $|v|<1$. Recall that
the $\beta=4+\de$, $0<\de<1/2$.
\begin{pro}\la{lindecay}
Let (\re{ro}), the Wiener condition (\re{W}), and the condition
(\re{zero4}) hold, $|v_1|\le\ti v<1$, and $X_0\in\cE$. Then
\\
i) The equation (\re{Avv}), with $A_1=A_{v_1,v_1}$, admits the unique solution
 $e^{A_1t}X_0:=X(t)\in C_b(\R, \cE)$
with the initial condition $X(0)=X_0$.
\\
ii) If $X_0\in{\cal Z}_{v_1}\cap{\cal E}_\beta$,
the following decay holds,
\be\la{frozenest}
\Vert e^{A_1t}X_0\Vert_{-2-\de}\le \fr{C(\ti v)}{(1+|t|)^{1+\de}}\Vert X_0\Vert_{\beta},\,\,\,t\in\R.
\ee
\end{pro}


Part i) follows by standard arguments using the positivity
(\re{H0vv}) of the Hamilton functional.
Part ii) will be proved in Sections \re{sFL}-\re{stdf}
developing general strategy \ci{IKV05,IKV06}.
Namely,
the equation (\re{Avv}) is a system of four
equations involving field components, $\Psi$ and $\Pi$ as well as
vector components, $Q$ and $P$. We apply Fourier-Laplace
transform, and express the field components in terms of the vector
components. Then we obtain a closed system for
the vector components alone and prove their decay. Finally, for
the field components we come to a wave equation with a right hand
side which has the established decay. This implies the
corresponding decay for the field components.

\setcounter{equation}{0}

\section{Solving the Linearized Equation}\la{sFL}
Here we
start the proof of Proposition
\re{lindecay}
solving the linearized equations
(\re{Avv}).
Let us apply the Laplace transform
\be\la{FL}
\Lam X=\ti
X(\lam)=\int_0^\infty e^{-\lam t}X(t)dt,~~~~~~~\Re\lam>0
\ee
to
(\re{Avv}). The integral converges in ${\cal E}$, since $\Vert X(t)\Vert_{\cal E}$
is bounded by Proposition \re{lindecay}, i). The
analyticity of $\ti X(\lam)$ and Paley-Wiener arguments should
provide the existence of a $\cE$-valued distribution $X(t)=(\Psi(t),\Pi(t),Q(t),P(t))$,
$t\in\R$, with a support in $[0,\infty)$. Formally,
\be\la{FLr}
\Lam^{-1}\ti X=X(t)=\fr1{2\pi}\int_\R e^{i\om t}\ti X(i\om+0)d\om,
~~~~~~~~t\in\R.
\ee
To prove the decay (\re{frozenest}), we have to
study the smoothness of $\ti X(i\om+0)$ at $\om\in\R$. After
 the Laplace transform the equation (\re{Avv}) becomes
$$
\lam\ti X(\lam)=A_1\ti X(\lam)+X_0,\,\,\,\Re\lam>0.
$$
In detail (for simplicity we write $v$ instead of $v_1$ in this section),
\be\la{eq1} \left.\ba{r}
\ti\Pi(y)+v\cdot\na\ti\Psi(y)-\lam\ti\Psi(y)=-\ti\Psi_0 (y)
\\
\\
\De\ti\Psi(y)+v\cdot\na\ti\Pi(y)+\ti
Q\cdot\na\rho(y)-\lam\ti\Pi(y)=-\Pi_0(y)
\\
\\
B_v\ti P-\lam\ti Q=-Q_0
\\
\\
-\langle\na\ti\Psi(y),\rho(y)\rangle+\langle\na\psi_{v}(y),\ti
Q\cdot\na\rho(y) \rangle-\lam\ti P=-P_0
\ea\right|~~~~~~~~~y\in\R^3. \ee
Let us consider the first two equations. In Fourier space  they become
\be\la{F1}
\left. \ba{rcl} \hat\Pi(k)-ivk\hat\Psi(k)-\lam\hat\Psi(k)&=&-\hat\Psi_0(k) \\\\
-k^2\hat\Psi(k)-(ivk+\lam)\hat\Pi(k)&=&-\hat\Pi_0(k)+i\ti Qk\hat\rho(k)
\ea\right|~~~~~~~~~k\in\R^3\5.
\ee
This implies
\be\la{FPS}
\hat\Psi=\fr1{\hat D}((ikv+\lam)\hat\Psi_0+\hat\Pi_0-ik\ti Q\hat\rho),
\ee
\be\la{FP}
\hat\Pi=\fr1{\hat D}(-k^2\hat\Psi_0+(ikv+\lam)\hat\Pi_0-i(ikv+\lam)k\ti Q\hat\rho),
\ee
where
\be\la{FD}
\hat D=\hat D(\lam)=k^2+(ikv+\lam)^2.
\ee
From now on we use the system of coordinates in $x$-space in which $v=(|v|,0,0)$,
hence $vk=|v|k_1$.
Substitute (\re{FPS}) to the 4-th equation of (\re{eq1}) and obtain
$$
\int\fr{ik}{\hat D}((ikv+\lam)\hat\Psi_0+\Pi_0-ik\ti Q\hat\rho)\ov{\hat\rho}dk
+\int k\hat\psi_vk\ti Q\ov{\hat\rho}dk-\lam\ti P=-P_0.
$$
Since $\hat\psi_v=-\hat\rho/(k^2-(kv)^2)$, we come to
$$
(K-H(\lam))\ti Q+\lam\ti P=P_0+\Phi(\lam).
$$
Here
\be\la{phidef}
\Phi(\lam)=\Phi(\Psi_0,\Pi_0)(\lam)
:=i\int\fr{k}{\hat D}((ikv+\lam)\hat\Psi_0
+\hat\Pi_0)\ov{\hat\rho}dk=i\langle
\fr{(ikv+\lam)\hat\Psi_0+\hat\Pi_0}{\hat D},
k\hat\rho
\rangle,
\ee
and $K$, $H(\lam)$ are $3\times3$-matrices with the matrix elements
\be\la{KHij}
K_{ij}=\int\fr{k_ik_j|\hat\rho(k)|^2dk}{\ds k^2-(|v|k_1)^2},\,\,\,H_{ij}(\lam)
=\int\fr{k_ik_j|\hat\rho(k)|^2dk}{k^2+(i|v|k_1+\lam)^2}.
\ee
The matrix $K$ is diagonal and positive definite since $\hat\rho(k)$  is
spherically symmetric and not identically zero by (\re{W}). The matrix $H$
is well defined for $\Re\lam>0$ since the denominator does not vanish.
The matrix $H$ is diagonal similarly to $K$. Indeed, if $i\ne j$, then at
least one of these indexes is not equal to one, and the integrand in (\re{KHij})
is odd with respect to the corresponding variable by (\re{ro}).
Finally the 3-rd and the 4-th equations of (\re{eq1}) become
\be\la{Mlam}
M(\lam)\left( \ba{c}\ti Q \\ \ti P \ea \right)=\left( \ba{c} Q_0 \\ P_0+\Phi(\lam)
\ea \right),\,\,{\rm where}\,\,M(\lam)=\left( \ba{cc} \lam I_3  & -B_{v} \\ -F(\lam) &
\lam I_3 \ea \right),\,\,\,F(\lam):=H(\lam)-K. \ee

\br{\rm
Note that
\be\la{philam}
\Phi(\lam)=\Phi(\Psi_0,\Pi_0)(\lam)=\Lam\langle W^1(t)(\Psi_0,\Pi_0),\na\rho\rangle,
\ee
where $W^1(t)$ is the first component of the dynamical group $W(t)$ defined below by
(\re{184st}) and $\Lam$ is the Laplace transform (\re{FL}).}
\er

Let us proceed to $x$-representation. We invert the matrix of the system (\re{F1})
and obtain
$$
\left( \ba{cc} -(ivk+\lam) & 1 \\ -k^2 & -(ivk+\lam)\ea
\right)^{-1}=[(ivk+\lam)^2+k^2]^{-1}\left( \ba{cc} -(ivk+\lam) & -1 \\ k^2 &
-(ivk+\lam) \ea \right).
$$
Taking the inverse Fourier transform, we obtain the corresponding fundamental
solution
\be\la{Green} G_{\lam}(y)=\left( \ba{cc} v\cdot\na-\lam & -1 \\ -\De &
v\cdot\na-\lam \ea\right)g_{\lam}(y),
\ee
where $g_{\lam}(y)$ is the unique tempered
fundamental solution of the determinant
\be \la{fso}
D=D(\lam)=-\De+(-v\cdot\na+\lam)^2.
\ee
Thus,
\be\la{dete} g_\lam(y)=F^{-1}_{k\to
y}\ds\fr{1}{k^2+(ivk+\lam)^2}= F^{-1}_{k\to y}\ds\fr{1}{k^2+
(i|v|k_1+\lam)^2},~~~y\in\R^3.
\ee
Note that the denominator does not vanish for
$\Re\lam>0$. This implies \bl\la{cres} The operator $G_{\lam}$ with the integral
kernel $G_{\lam}(y-y')$ is continuous operator $H^1(\R^3)\oplus L^2(R^3)\to
H^2(\R^3)\oplus H^1(R^3)$  for $\Re\lam>0$. \el
From (\re{FPS}) and (\re{FP})
we obtain the convolution representation \be\la{Psi} \left.\ba{rcl}
\Psi&=&-(v\cdot\na-\lam)g_{\lam}*\Psi_0+g_{\lam}*\Pi_0+(g_{\lam}*\na\rho)\cdot Q \\\\
\Pi&=&\De g_{\lam}*\Psi_0-
(v\cdot\na-\lam)g_{\lam}*\Pi_0-(v\cdot\na-\lam)(g_{\lam}*\na\rho)\cdot Q \ea\right|
\ee


\noindent Let us compute $g_{\lam}(y)$ explicitly. First consider the
case $v=0$. The fundamental solution of the operator $-\De+\lam^2$ is \be\la{fus}
g_{\lam}(y)=\fr{e^{-\lam|y|}}{4\pi|y|}. \ee
Thus, in the case $v=0$ we have
$$
G_{\lam}(y-y')=\left( \ba{cc} -\lam & -1 \\ -\De & -\lam \ea \right)
\fr{e^{-\lam|y-y'|}}{4\pi|y-y'|}.
$$
For general $v=(|v|,0,0)$ with $|v|<1$ the denominator in (\re{dete}), which is the
Fourier symbol of $D$, reads
\beqn \hat D(k)&=& k^2+(i|v|k_1+\lam)^2 = (1-v^2)k_1^2+k_2^2+k_3^2+ 2i|v|k_1\lam+\lam^2
\nonumber\\ &=& (1-v^2)(k_1+\fr{i|v|\lam}{1-v^2})^2+k_2^2+k_3^2+ \ka^2, \la{Dhat}
\eeqn where \be\la{tikappa}
\ka^2=\fr{v^2\lam^2}{1-v^2}+\lam^2=\fr{\lam^2}{1-v^2}. \ee Therefore, setting
$\ga:=1/\sqrt{1-v^2}$, we have \be\la{kappa} \ka=\ga\lam. \ee
Return to $x$-space: \be\la{13.10'}
D=-\fr{1}{\ga^2}(\na_1+\ga\ka_1)^2- \na_2^2-\na_3^2+\ka^2, ~~~~~~~~ \ka_1:=\ga|v|\lam
\ee Define $\ti y_1:=\ga y_1$ and $\ti\na_1:=\pa/\pa \ti y_1$. Then \be\la{Dy}
D=-(\ti\na_1+\ka_1)^2-\na_2^2-\na_3^2+\ka^2. \ee Thus, its fundamental solution is
\be\la{glam} g_{\lam}(y)=\fr{e^{-\ka|\ti y|-\ka_1\ti y_1}}{4\pi|\ti y|}, \,\,\,\ti
y:=(\ga y_1,y_2,y_3).\ee
By (\re{kappa}), (\re{13.10'}) we obtain
\be\la{kak} 0<\Re\ka_1<\Re\ka,~~~~~~\Re\lam>0. \ee
Let us state the result which
we have got above.
\bl
\la{cac} i) The operator $D=D(\lam)$ is invertible in
$L^2(\R^3)$ for $\Re\lam>0$ and its fundamental solution (\re{glam}) decays
exponentially in $y$.

ii) The formulas (\re{glam}) and (\re{kappa}), (\re{13.10'}) imply
that
the distribution
 $g_\lam(\cdot)$ admits an analytic continuation in the papameter
$\lam$ from the domain
 $\Re\lam>0$
to the entire
complex plane $\C$. \el

\bl\la{cmf} The matrix function $M(\lam)$ (respectively, $M^{-1}(\lam)$) admits an
analytic (respectively meromorphic) continuation from the domain
 $\Re\lam>0$
to the entire
complex plane $\C$. \el
\pru From the first equation of (\re{Psi}) and the last equation of (\re{eq1})
it follows that \be\la{Hjjx}
H_{jj}(\lam)=\langle g_\lam*\pa_j\rho,\pa_j\rho\rangle
\ee
and thus, by (\re{glam}),
$$
|H_{jj}(\lam)|=|\langle g_\lam*\pa_j\rho,\pa_j\rho\rangle|\le\int\fr{C}{|x-y|}
\pa_j\rho(x)\pa_j\rho(y)dxdy<\infty.
$$
By (\re{KHij}), (\re{Mlam}) this implies
\be\la{Flambound}
\sup\limits_{\Re\lam\ge0}|F(\lam)|<\infty.
\ee
The analytic continuation of  $M(\lam)$
exists by the expressions (\re{Hjjx}) and Lemma \re{cac} ii), since the function
$\rho(x)$ is compactly supported by  (\re{ro}).  The inverse matrix is then
meromorphic since it exists for large $\Re\lam$. The latter follows from (\re{Mlam})
since $H(\lam)\to 0$, $\Re\lam\to\infty$, by  (\re{KHij}).\hfill $\Box$

\setcounter{equation}{0}

\section{Regularity in Continuous Spectrum}
\la{srcs}

\bp\la{creg} The matrix $M^{-1}(i\om)$ is analytic in $\om\in\R\setminus\{0\}$. \ep

\noindent\pru It suffices to prove that the limit matrix
 $M(i\om):=M(i\om+0)$ is invertible for
$\om\ne 0$, $\om\in\R$ if $\rho$ satisfies the Wiener condition (\re{W}), and
$|v|<1$. Let $|\om|>0$.
One has
\beqn\la{detM}
{\rm det}\,M(i\om)={\rm det}\,\left(
\ba{ll}
i\om I_3  & -B_v \\
-F(i\om) & i\om I_3 \ea \right)=-(\om^2+\nu^3f_1)(\om^2+\nu f)^2,~~~\om\in\R, \eeqn
where $F(i\om):=F(i\om+0)$, $f_1:=F_{11}(i\om)$, and $f:=F_{22}(i\om)=F_{33}(i\om)$.
The invertibility  of $M(i\om)$ follows from  (\re{detM}) by the following lemma,
whose proof is based on the Sokhotsky-Plemelj formula, see \ci[Chapter VII, formula
(58)]{Vai}. \bl\la{lW} If (\re{W}) holds, then for $\om\in\R$ the imaginary part of
the matrix $\ds\fr{\om} {|\om|}F(i\om)$ is negative definite, i.e.
$\ds\fr{\om}{|\om|}\Im F_{jj}(i\om)<0$, $j=1,2,3$. \el \pru Since
$F(i\om)=H(i\om+0)-K$, where the matrix $K$ is real, we will consider only the matrix
$H(i\om+0)$. For $\ve>0$ we have \be\la{Hjjlim}
H_{jj}(i\om+\ve)=\int\fr{k_j^2|\hat\rho(k)|^2dk}{k^2-(|v|k_1+\om-i\ve)^2},\,\,\,j=1,2,3.
\ee Consider the denominator
$$
\hat D(i\om+\ve,k)=k^2-(|v|k_1+\om-i\ve)^2.
$$
$\hat D(i\om,k)=0$  on the ellipsoid $T_{\om}$ if $|\om|>0$, where
$$
T_{\om}=\{k:(\nu k_1-\fr{|v|\om}{\nu})^2+k_2^2+k_3^2=\fr{\om^2}{\nu^2}\},
$$
here $\nu=\sqrt{1-v^2}$.
From the Sokhotsky-Plemelj formula for $C^1$-functions it follows that
\be\la{ImHjj}
\Im H_{jj}(i\om+0)=-\fr{\om}{|\om|}\pi\int_{T_{\om}}\fr{k_j^2|\hat\rho(k)|^2} {|\na\hat D(i\om,k)|}dS,
\ee
where $dS$ is the element of the surface area. This immediately implies the
statement of the Lemma since the integrand in (\re{ImHjj}) is positive by the Wiener condition (\re{W}).
 This completes the proofs of the lemma
and the Proposition \re{creg}. \hfill$\Box$

\br {\rm The proof of Lemma \re{lW} is the unique point in the
paper where the Wiener condition is indispensable.}
\er

\setcounter{equation}{0}
\section{Time Decay of the Vector Components}
\la{stdvc}
Here we prove the decay (\re{frozenest})
for the vector components $Q(t)$ and $P(t)$
of the solution $e^{A_1t}X_0$.
Formula (\re{Mlam}) expresses the Laplace transforms
$\ti Q(\lam),\ti P(\lam)$. Hence, the components are given by the integral
\be\la{QP1ii}
\left( \ba{c} Q(t) \\ P(t) \ea \right)=\ds\fr
1{2\pi}\int e^{i\om t}M^{-1}(i\om)\left( \ba{c} Q_0 \\ P_0+\Phi(i\om)\ea
\right)d\om.
\ee

Let us recall that in Proposition \re{lindecay} ii) we assume that
\be\la{beta3}
X_0\in{\cal Z}_v\cap \cE_{\beta},\,\,\,\beta=4+\de,\,\,\,0<\de<1/2.
\ee
\bt\la{171}
The functions $Q(t)$, $P(t)$ are continuous for $t\ge 0$,
 and
\be\la{decQP}
|Q(t)|+|P(t)|\le \ds\fr {C(\rho, \ti v)}{(1+|t|)^{1+\de}}\Vert X_0\Vert_{\beta}, ~~~~~~~t\ge 0. \ee
\et
\pru Note that the Proposition \re{creg} alone is not sufficient for the proof of the
convergence and decay of the integral (\re{QP1ii}). We need an additional information
about the regularity of the matrix $M^{-1}(i\om)$ at its singular point  $\om=0$, and
some bounds at $|\om|\to\infty$.

Let us split the integral (\re{QP1ii}) in two terms using the partition of unity
$\zeta_1(\om)+\zeta_2(\om)=1$, $\om\in\R$:
\be\la{QP1i3}
\left( \ba{c}
Q(t) \\
P(t)
\ea \right)=\ds\fr 1{2\pi}\int e^{i\om t}(\zeta_1(\om)+\zeta_2(\om))
\left( \ba{c}
\ti Q(i\om) \\
\ti P(i\om)
\ea \right)
d\om = \left( \ba{c}
Q_1(t) \\
P_1(t)
\ea \right)+\left( \ba{c}
Q_2(t) \\
P_2(t)
\ea \right),
\ee
where the functions $\zeta_k(\om)\in C^\infty(\R)$ are supported by
\be\la{zsup}
\supp \zeta_1\subset\{\om\in\R:|\om|<r+1\},\,\,\,
\supp\zeta_2\subset\{\om\in\R:|\om|>r\},\ee
where $r$ is introduced below in Lemma \re{162}.
We prove the decay (\re{decQP}) for $(Q_1,P_1)$
and $(Q_2,P_2)$ in Propositions \re{PQtdec} and \re{Q2P2} respectively.
\begin{pro}\la{Q2P2} The functions $Q_2(t)$, $P_2(t)$ are continuous
for $t\ge 0$, and
\be\la{I2dec}
|Q_2(t)|+|P_2(t)|\le\ds\fr {C(\rho, \ti v)}{(1+|t|)^{3+\de}}\Vert X_0
\Vert_{\beta}.
\ee
\end{pro}
\pru First we study the asymptotic behavior of $M^{-1}(\lam )$ at infinity.
Let us recall that
$M^{-1}(\lam )$ was originally defined for $\Re\lam >0,$ but it admits
a meromorphic
continuation to $\C$ (see Lemma \ref{cmf}).
\bl \la{162}
There exist a matrix $R_{0}$ and a matrix-function $R_{1}(\om )$,  such that
\be\la{Minf}
M^{-1}(i\om)=\fr {R_0}\om +R_1(\om ),~~~|\om|>r>0,~~~~~~~~~\om \in \R,
\ee
where, for every $k=0,1,2,...,$
\be\label{min}
|\pa_\om^k R_1(\om )|\leq \fr{C_k}{|\om |^2}, ~~~~~~~~~~~~|\om|>r>0,~~~~~~~~~\om \in \R,
\ee
$r$ is sufficiently large.
\el
\pru The statement follows from the expilicit formulas (\re{invinv})
 to (\re{inv22}) for the inverse matrix $M^{-1}(i\om)$ and from the bound (\re{Flambound}).
\hfill $\Box$

\medskip

 Further, (\re{QP1ii}) implies that
$$
\left( \ba{c}
Q_2(t) \\
P_2(t)
\ea \right)=\fr 1{2\pi}\int e^{i\om t}\zeta_2(\om)M^{-1}(i\om)
\left[
\left(\ba{c}
Q_0 \\
P_0
\ea \right)+\left( \ba{c}
0 \\
\Phi(i\om)
\ea \right)
\right]d\om
$$
\be\la{long}
=s(t)\left(
\ba{c}
Q_0 \\ P_0
\ea
\right)+s*\left(
\ba{c}
0 \\ f
\ea
\right),
\ee
where (see(\re{FLr}))
$$
s(t):=\Lam^{-1}\left[\zeta_2(\om)M^{-1}(i\om)\right]
$$
and
\be\la{fphi}
f(t):=\Lam^{-1}\Phi(i\om)=\langle W^1(t)(\Psi_0,\Pi_0),\na\rho\rangle,
\ee
since $\Phi$ is given by (\re{phidef}) and (\re{philam}).
Note that $s(t)$ is continuous
for $t\ge 0$, and
\be\la{sNN}
|s(t)|={\cal O}(t^{-N}),\,\,\, t\to\infty,\,\,\,\forall\,N>0.
\ee
by (\re{Minf})-(\re{min}).
On the other hand,
recall that, under the conditions of Proposition \re{lindecay} and Theorem \re{main}, $(\Psi_0,\Pi_0)\in \cF_\beta$ with
$\beta=4+\de$
where $\de>0$.
Then we
obtain that

\be\la{fdec}
|f(t)|\le\ds\fr {C(\rho, \ti v)}{(1+|t|)^{3+\de}}\Vert X_0\Vert_{\beta}
\ee
by Lemma \re{lemma} below with $\al=\beta$.
Hence,  (\re{I2dec}) follows from (\re{long}) by
(\re{sNN}) and (\re{fdec}).\hfill $\Box$

\bigskip

\noindent Now let us prove the decay for $Q_1(t)$ and $P_1(t)$. In
this case
the proof will rely
substantially on the symplectic orthogonality conditions.
Namely, (\re{beta3}) implies that
\be\la{ort16}
\Om(X_0,\tau_j)=0,\,\,\,j=1\dots6.
\ee
\bp\la{PQtdec} The functions $Q_1(t)$ and $P_1(t)$ are continuous
for $t\ge 0$, and
\be\la{PQdec}
|Q_1(t)|+|P_1(t)|\le\fr{C(\rho,\ti v)\Vert
X_0\Vert_{\beta}}{(1+t)^{1+\de}},
\,\,\,t\ge0.
\ee
\ep
\pru First, we obtain  the formulas for the Fourier transforms
$\ti Q(i\om)$ and $\ti P(i\om)$.

\bl\la{Minv}
The matrix  $M^{-1}(i\om)$ can be represented as follows,
\be\la{calLom}
L(\om):=M^{-1}(\om)=\left(
\ba{cc}
\fr{\ts 1}{\ds\om}{\cal L}_{11} & \fr{\ts 1}{\ds\om^2}{\cal L}_{12} \\
{\cal L}_{21} & \,\,\,\fr{\ts 1}{\ds\om}{\cal L}_{22}
\ea
\right),
\ee
where ${\cal L}_{ij}(\om)$, $i,j=1,2$ are smooth diagonal $3\times3$-matrices,
${\cal L}_{ij}(\om)\in C^{\infty}(-r-1;r+1)$, and
\be\la{m11m12}
{\cal L}_{11}=i{\cal L}_{12}B_v^{-1}.
\ee
\el
For proof see Appendix A.
Now (\re{Mlam}) implies that the vector components are given by
\be\la{tiQiom}
\ti Q(i\om)=\fr{1}{\om}{\cal L}_{11}(\om)Q_0+\fr{1}{\om^2}{\cal L}_{12}(\om)(P_0+\Phi(i\om)),
\ee
\be\la{tiPiom}
\ti P(i\om)={\cal L}_{21}(\om)Q_0+\fr{1}{\om}{\cal L}_{22}(\om)(P_0+\Phi(i\om)),
\ee

Next we calculate the symplectic orthogonality conditions (\re{ort16}).
\bl\la{sort16} The symplectic orthogonality conditions (\re{ort16}) read
\be\la{ort123}
P_0+\Phi(0)=0\,\,\,\,{\rm and}\,\,\,\,
B_v^{-1}Q_0+\Phi'(0)=0.
\ee
\el
For proof see Appendix B.

\smallskip
Now we can prove Proposition
\re{PQtdec}.

\noindent{\it Step i)} Let us prove (\re{PQdec}) for $P_1(t)$ relying on the representation
(\re{tiPiom}). Namely, (\re{QP1i3}) and (\re{tiPiom}) imply
$$
P_1(t)=\Lam^{-1}\zeta_1(\om){\cal L}_{21}(\om)Q_0+\Lam^{-1}\zeta_1(\om){\cal L}_{22}(\om)
\fr{P_0+\Phi(i\om)}{\om}=P_1'(t)+P_1''(t).
$$
The first term $P'_1(t)$ decays like $Ct^{-\infty}\Vert X_0\Vert_{\beta}$ by
Lemma \re{Minv}. The second term admits the convolution representation $P_1''(t)
=\Lam^{-1}\zeta_1{\cal L}_{22}*g(t)$, where
$$
g(t):=\Lam^{-1}\fr{P_0+\Phi(i\om)}{\om}.
$$
Now we use the symplectic orthogonality conditions (\re{ort123}) and obtain
\be\la{goft}
g(t)=\Lam^{-1}\fr{\Phi(i\om)-\Phi(0)}{\om}=i\int\limits^{t}_{\infty}f(s)ds.
\ee
Therefore, $P_1''(t)$ decays like $Ct^{-(2+\de)}\Vert X_0\Vert_{\beta}$ for $t\ge0$,
since by (\re{fdec})
\be\la{goftdec}
|g(t)|\le C(\rho,\ti v)(1+t)^{-(2+\de)}\Vert X_0\Vert_{\beta},\,\,\,t\ge0.
\ee

\medskip

\noindent{\it Step ii)} Now let us prove (\re{PQdec}) for $Q_1(t)$. By (\re{tiQiom}),
(\re{m11m12}), and the symplectic orthogonality conditions (\re{ort123}),
$$
\ti Q(i\om)=\fr{{\cal L}_{12}}{\om}\left(iB_v^{-1}Q_0+\fr{P_0+\Phi(i\om)}{\om}\right)=
\fr{{\cal L}_{12}}{\om}\left(iB_v^{-1}Q_0+\fr{\Phi(i\om)-\Phi(0)}{\om}\right)=
$$
$$
\fr{{\cal L}_{12}}{\om}(iB_v^{-1}Q_0+\ti g(\om))=
{\cal L}_{12}\fr{iB_v^{-1}Q_0+\ti g(0)+\ti g(i\om)-\ti g(0)}{\om}={\cal L}_{12}\fr{\ti
g(i\om)-\ti g(0)}{\om},
$$
since $iB_v^{-1}Q_0+\ti g(0)=0$ by the symplectic orthogonality conditions (\re{ort123}),
because $\ti g(0)=i\Phi'(0)$. Thus, $Q_1(t)=\Lam^{-1}\zeta_1(\om){\cal L}_{12}*h(t)$ by
(\re{QP1i3}), where
$$
h(t):=\Lam^{-1}\fr{\ti g(i\om)-\ti g(0)}{\om}=i\int\limits^{t}_{\infty}g(s)ds,
$$
similarly to (\re{goft}). This integral decays like $Ct^{-1-\de}\Vert X_0\Vert_{\beta}$
for $t\ge0$ by (\re{goftdec}), hence (\re{PQdec}) for $Q_1(t)$ is proved.
The proof of Proposition \re{PQtdec} and Theorem \re{171} is complete.\hfill $\Box$

\setcounter{equation}{0}
\section{Time Decay of Fields}\la{stdf}

Here we construct the field components $\Psi(x,t),\Pi(x,t)$ of the
solution $X(t)$
and prove their decay corresponding to (\re{frozenest}).
Let us denote
$
F(t)=(\Psi(\cdot,t),\Pi(\cdot,t)).
$
We will construct the fields solving the first two equations of
 (\re{Avv}), where $A$ is given by (\re{AA}).
These two equations have the form
\be\la{linf}
\dot F(t)=\left(
\ba{ll}
v\cdot\na & 1 \\
\De & v\cdot\na
\ea
\right)F+\left(
\ba{l}
0 \\
\na\rho\cdot Q(t)
\ea
\right).
\ee
By
Theorem \re{171}
 we know that $Q(t)$ is continuous and
\be\la{linb}
|Q(t)|\le \ds\fr{ C(\rho, \ti v)\Vert X_0\Vert_{\beta} }{ (1+t)^{1+\de} },~~~~
t\ge 0.
\ee
Hence, the Proposition \re{lindecay}
is reduced now to the following
\begin{pro} \la{pfi}
i) Let a function $Q(t)\in C([0,\infty);\R^3)$,
and $F_0\in\cF$.
Then
the equation (\re{linf}) admits a unique
solution $F(t)\in C[0,\infty;\cF)$ with the initial condition
$F(0)=F_0$.
\\
ii) If $X_0=(F_0;Q_0,P_0)\in{\cal E}_{\beta}$ and the decay (\re{linb}) holds, the
corresponding fields also decay
uniformly in $v$:
\be\la{lins}
\Vert F(t)\Vert_{-2-\de}\le \ds\fr{C(\rho,\ti v)\Vert X_0\Vert_{\beta}}{(1+t)^{1+\de}}
,~~~~t\ge 0,
\ee
for $|v|\le \ti v$ with any $\ti v\in (0;1)$.
\end{pro}
{\bf Proof }  The statement i) follows from the Duhamel representation
\be\la{Duh} F(t)=W(t)F_0+\left[\int_0^tW(t-s)\left( \ba{l} 0 \\ \na\rho\cdot Q(s) \ea
\right)ds \right],~~~~~~t\ge 0,
\ee
where $W(t)$ is the dynamical group of the modified wave equation
\be\la{184st} \dot F(t)=\left( \ba{cc} v\cdot\na & 1 \\ \De & v\cdot\na \ea \right)F(t).
\ee
The group $W(t)$ can be expressed through the group $W_0(t)$ of the wave  equation
\be\la{KG} \dot \Phi(t)=\left( ~~ \ba{cc} 0 & 1 \\ \De ~&~ 0 \ea ~~ \right)\Phi(t).
\ee
Namely, the problem (\re{KG}) corresponds to (\re{184st}), when $v=0$, and it is easy to see that
\be\la{shifted} [W(t)F(0)](x)=[W_0(t)F(0)](x+vt),~~~~x\in\R^3,~~t\in\R.
\ee
The identity (\re{shifted}) implies the energy  conservation law for the group $W(t)$: for
$(\Psi(\cdot,t),\Pi(\cdot,t))=W(t)F(0)$ we have
$$
\int\left(|\Pi(x,t)-v\cdot\na\Psi(x,t)|^2+|\na\Psi(x,t)|^2\right)dx=\co, ~~~~~~t\in\R.
$$
In particular, this gives by (\re{FFF})
\begin{equation}
\Vert W(t)F_{0}\Vert _{\mathcal{F}}\leq C(\ti{v})\Vert F_{0}\Vert _{\mathcal{F}},\,\,\,t\in\R.
\label{E}
\end{equation}
This estimate and (\ref{Duh}) imply the statement i).

\medskip

\noindent Let us proceed to the statement ii).

\bl\label{lemma}
For $\ti{v}<1$ and $F_{0}\in \mathcal{F}_{\al}$, $\al>1$, the following decay holds,
\begin{equation}
\Vert W(t)F_{0}\Vert _{-\al}\leq \frac{C(\al ,\ti{v})}{(1+t)^{\al-1} }\Vert F_{0}\Vert _{\al },
~~~~~~t\geq 0,  \label{dede}
\end{equation}
for the dynamical group $W(t)$ corresponding to the modified wave equation (\ref{184st}) with
$|v|<\ti{v}$.
\el
\pru For the case $v=0$ the proof is provided in \ci{Kop}. For a nonzero $v$ with $|v|<\ti{v}$
the proof is similar, we provide it for convenience.

We should estimate $\Vert W(t)F_0\Vert_{-\al}$ for large $t>0$.
Set $\ve=(1-\ti v)/2$. For
an arbitrary sufficiently large $ t\ge 1$ let us split the initial function $F_0$ in two terms,
 $F_0=F_{0,t}'+F_{0,t}''$ such that
\be\la{FFna}
\ba{l}
\Vert F_{0,t}'\Vert_\al +\Vert F_{0,t}''\Vert_\al
\le C \Vert F_0\Vert_\al,~~~~~~~t\gg 1,
\ea
\ee
where $C$ does not depend on $t$,
and
\be\la{F'a}
F_{0,t}'(x)=0,\,\,\,|x|>\ve t,\,\,\,\,F_{0,t}''(x)=0,\,\,\,|x|<\ve t-1.
\ee
For an arbitrary $f\in\dot H^1$ and $\al>1$ one has $\Br f\Br_{-\al}\le C\Vert f\Vert_{\dot H^1}$,
see \ci{Kop}, formula (2.9). Now the estimate for $W(t)F_{0,t}''$
 follows by (\re{E}), (\re{F'a})  and  (\re{FFna}) :
\be\la{cherepa}
\Vert W(t)F_{0,t}''\Vert_{-\al}\le C\Vert W(t)F_{0,t}''\Vert_{\cal F}
\le C\Vert F_{0,t}''\Vert_{\cal F}
\le
C_1(\ti v)\Vert F_{0,t}''\Vert_{\al}(1+|t|)^{-\al}\le
C_2(\ti v)\Vert F_0\Vert_{\al}(1+|t|)^{-\al},\,\,\,t\ge 1.
\ee

It remains to estimate $W(t)F_{0,t}^{\prime}$. First note that
\be\la{W0sup}
W_0(t)F_{0,t}^{\prime}(x)=0\,\,\,{\rm for}\,\,\,|x|<(1-\ve)t
\ee
by the strong Huygen's principle for the group $W_0(t)$. The principle reads
\be\la{Wrea}
W_0(x-y,t)=0,~~~~|x-y|\ne t,
\ee
where $W_0(z,t)$ is the integral (distribution) matrix kernel of the operator $W_0(t)$.
Further, from (\re{W0sup}) it follows that
$$
[W(t)F_{0,t}'](x)=0\,\,\,{\rm for}\,\,\,|x|<\ve t
$$
by (\re{shifted}) and since $|v|<\ti v=1-2\ve$.

For an arbitrary $f\in\dot H^1$ such that $f(x)=0$ in the region $\{|x|<\ve t\}$, one has
$\Br f\Br_{-\al}\le C(\ti v)t^{-\al+1}\Vert f\Vert_{\dot H^1}$, see \ci{Kop}, the proof of
Proposition 2.1. Applying to $f=W(t)F'_{0,t}$ we obtain by (\re{E}) that,
$$
\Vert W(t)F'_{0,t}\Vert_{-\al}\le C(\ti v)t^{-\al+1}\Vert W(t)F'_{0,t}\Vert_{\cal F}
\le Ct^{-\al+1}\Vert F'_{0,t}\Vert_{\cal F}\le Ct^{-\al+1}\Vert F'_{0,t}\Vert_{{\cal F}_\al}
\le Ct^{-\al+1}\Vert F_{0}\Vert_{{\cal F}_\al}.
$$
The proof is complete. \hfill $\Box$

\medskip

\noindent
Now the statement ii) of Proposition \re{pfi} follows from the Lemma \re{lemma} and the Duhamel
representation (\ref{Duh}). Indeed,
$$
\Vert W(t)F_0\Vert_{-2-\de}\le Ct^{-1-\de}\Vert F_0\Vert_{2+\de}\le Ct^{-1-\de}\Vert X_0
\Vert_{2+\de}\le Ct^{-1-\de}\Vert X_0\Vert_{\beta}
$$
by Lemma \re{lemma} with $\alpha=2+\de$. Further,
$$
\Vert\int\limits_0^t W(t-s)\left(
\ba{c} 0 \\ \na\rho\cdot Q(s) \ea
\right)ds\Vert_{-2-\de}\le
C\int\limits_0^t\fr{\Vert(0,\na\rho\cdot Q(s))\Vert_{2+\de}ds}{(1+(t-s))^{1+\de}}
$$
$$
\le
C'\int\limits_0^t\fr{|Q(s)|ds}{(1+(t-s))^{1+\de}}
\le C''\Vert X_0\Vert_{\beta}\int\limits_0^t\fr{ds}{(1+(t-s))^{1+\de}(1+s)^{1+\de}}
$$
by Lemma \re{lemma} with $\alpha=2+\de$, regularity properties of $\rho$, and (\re{linb}).
The last integral decays like $(1+t)^{-1-\de}$ by a well known result on decay of a convolution.
\hfill$\Box$

The proof of  Proposition \re{lindecay} is complete.\hfill$\Box$

\setcounter{equation}{0}

\section{Frozen Form of Transversal Dynamics}
In next four sections we will prove
the transversal decay (\re{Zdec})
relying on the Proposition \re{lindecay}.
First,  let us fix an arbitrary $t_1\in [0,t_*)$, and
rewrite the equation (\re{reduced}) in a ``frozen form''
\be\la{froz}
\dot Z(t)=A_1Z(t)+(A(t)-A_1)Z(t)+\ti N(t),\,\,\,~~~~0\le t<t_*,
\ee
where $A_1=A_{v(t_1),v(t_1)}$ and
$$
A(t)-A_1=\left(
\ba{cccc}
[w(t)\!-\!v(t_1)]\cdot \na & 0 & 0 & 0 \\
0 & [w(t)\!-\!v(t_1)]\cdot \na & 0 & 0 \\
0 & 0 & 0 & B_{v(t)}\!-\!B_{v(t_1)} \\
0 & 0 &
\langle\na(\psi_{v(t)}\!-\!\psi_{v(t_1)}),\cdot\na\rho\rangle & 0
\ea \right).
$$
The next trick is important since it allows us to kill the ``bad terms''
 $[w(t)\!-\!v(t_1)]\cdot \na$ in the operator $A(t)-A_1$.
\begin{definition}\la{d71}
Let us change the  variables $(y,t)\mapsto (y_1,t)=(y+d_1(t),t)$
where
\be\la{dd1}
d_1(t):=\int_{t_1}^t(w(s)-v(t_1))ds, ~~~~0\le t\le t_1.
\ee
\end{definition}
Next define
\beqn\la{Z1}
Z_1(t)&=&(\Psi_1(y_1,t),\Pi_1(y_1,t),Q(t),P(t)):=
(\Psi(y,t),\Pi(y,t),Q(t),P(t))\nonumber\\
&=&
(\Psi(y_1-d_1(t),t),\Pi(y_1-d_1(t),t),Q(t),P(t)).
\eeqn
Then we obtain the final form of the
``frozen equation'' for the transversal dynamics
\be\la{redy1}
\dot Z_1(t)=A_1Z_1(t)+B_1(t)Z_1(t)+N_1(t),\,\,\,0\le t\le t_1,
\ee
where $N_1(t)=\ti N(t)$ from the basic equation (\re{reduced}) expressed in terms of $y=y_1-d_1(t)$,
and
$$
B_1(t)=\left(
\ba{cccc}
0 & 0 & 0 & 0 \\
0 & 0 & 0 & 0 \\
0 & 0 & 0 & B_{v(t)}\!-\!B_{v(t_1)} \\
0 & 0 & \langle\na(\psi_{v(t)}\!-\!\psi_{v(t_1)}),\cdot\na\rho\rangle & 0
\ea
\right).
$$
At the end of this section, we will derive appropriate bounds for the
``remainder terms'' $B_1(t)Z_1(t)$ and $N_1(t)$ in (\re{redy1}).
First, note that we have by Lemma \re{mod},
\be\la{BB1}
|B_{v(t)}-B_{v(t_1)}|\le|\int_{t_1}^t\dot v(s)\cdot\na_vB_{v(s)}ds|
\le C\int_{t}^{t_1}\Vert Z(s)\Vert_{-\beta}^2ds.
\ee
Similarly,
\be\la{psipsi1}
|\langle\na(\psi_{v(t)}-\psi_{v(t_1)}),\cdot\na\rho\rangle|\le
 C\int_t^{t_1}\Vert Z(s)\Vert_{-\beta}^2ds.
\ee
Let us recall the following
well-known inequality:
for any $\al\in\R$
\be\la{pitre}
(1+|y+x|)^{\alpha}\le(1+|y|)^{\alpha}(1+|x|)^{|\alpha|},
\,\,\,~~~~~~x,y\in\R^3.
\ee
\begin{lemma}\la{dest}{\rm \ci[Lemma 7.2]{IKV05}}.
For $(\Psi,\Pi,Q,P)\in{\cal E}_{\alpha}$ with any $\alpha\in\R$
the following estimate holds:
\be\la{shiftest}
\Vert(\Psi(y_1-d_1),\Pi(y_1-d_1),Q,P)\Vert_{\alpha}\le
\Vert(\Psi,\Pi,Q,P)\Vert_{\alpha}(1+|d_1|)^{|\alpha|}~,\,\,\,~~~~~~d_1\in\R^3.
\ee
\end{lemma}

\begin{cor}\la{cor1}
The following  bound holds
\be\la{N1est}
\Vert N_1(t)\Vert_{\beta}\le\Vert Z_1(t)\Vert^2_{-\beta}
(1+|d_1(t)|)^{3\beta}~,~~~0\le t\le t_1.
\ee
\end{cor}
Indeed, applying the previous lemma twice, once for $\al=\beta$ and once
for $\al=-\beta$, we obtain from (\re{redN}) that
$$
\Vert N_1(t)\Vert_{\beta}\le(1+|d_1(t)|)^{\beta}
\Vert \ti N(t,Z(t))\Vert_{\beta}\le
(1+|d_1(t)|)^{\beta}\Vert Z\Vert^2_{-\beta}\le(1+|d_1(t)|)^{3\beta}\Vert
Z_1(t)\Vert^2_{-\beta}~.
$$
\begin{cor}\la{cor2}
The following bound holds
\be\la{B1Z1est}
\Vert B_1(t)Z_1(t)\Vert_{\beta}\le C
\Vert Z_1(t)\Vert_{-\beta}\int_t^{t_1}(1+|d_1(\tau)|)^{2\beta}
\Vert Z_1(\tau)\Vert^2_{-\beta}
d\tau~,~~~0\le t\le t_1.
\ee
\end{cor}
For proof we apply Lemma \re{dest} with $\al=-\beta$ to (\re{BB1}) and (\re{psipsi1})
and use the fact that $B_1(t)Z_1(t)$ depends only on the finite-dimensional
components of $Z_1(t)$.


\setcounter{equation}{0}

\section{Integral Inequality}

Recall that $0<\de<1/2$.
The equation (\re{redy1}) can be written in the integral form:
\be\la{Z1duh}
Z_1(t)=e^{A_1t}Z_1(0)+\int_0^te^{A_1(t-s)}[B_1Z_1(s)+N_1(s)]ds,\,\,\,
0\le t\le t_1.
\ee
We apply the symplectic orthogonal
projection $\bP_1:=\bP_{v(t_1)}$ to both sides, and get
$$
\bP_1Z_1(t)=e^{A_1t}\bP_1Z_1(0)+\int_0^te^{A_1(t-s)}\bP_1[B_1Z_1(s)+N_1(s)]ds.
$$
We have used here that $\bP_1$ commutes
with
the group $e^{A_1t}$ since the space $\cZ_1:=\bP_1\cE$ is invariant
with respect to $e^{A_1t}$, see Remark \re{rZ}.
Applying (\re{frozenest}) we obtain that
\be\la{bPZ}
\Vert\bP_1Z_1(t)\Vert_{-2-\de}\le\fr{C}{(1+t)^{1+\de}}\Vert Z_1(0)\Vert_{\beta}+C\int_0^t
\fr1
{(1+|t-s|)^{1+\de}}\Vert B_1Z_1(s)+N_1(s)\Vert_{\beta}ds,
\ee
since the operator $\bP_1$ is continuous in ${\cal E}_\beta$.
Hence,  from (\re{bPZ}) and (\re{N1est}),
(\re{B1Z1est}) we obtain that
\beqn\la{duhest}
\!\!\!\!\!\!\!\!\!\!\!\!\!\!\!\!\!\!&&\!\!\!\!\!\! \Vert
\bP_1Z_1(t)\Vert_{-2-\de}\le\fr{C}{(1+t)^{1+\de}}\Vert
Z_1(0)\Vert_{\beta}
\nonumber\\
\!\!\!\!\!\!\!\!\!\!\!\!\!\!\!\!\!\!&&\!\!\!\!\!\! +\,C(\ov
d_1)\int_0^t\fr1{(1+|t-s|)^{1+\de}}\left[\Vert
Z_1(s)\Vert_{-\beta} \int_s^{t_1}\Vert
Z_1(\tau)\Vert^2_{-\beta}d\tau+ \Vert
Z_1(s)\Vert^2_{-\beta}\right]ds,\,\,\,0\le t\le t_1, \eeqn where
$\ov d_1:=\sup_{0\le t\le t_1} |d_1(t)| $. Since
$\Vert Z_1(t)\Vert_{\pm\beta}\le C(\ov d_1) \Vert
Z(t)\Vert_{\pm\beta}$ by Lemma \re{dest}, we can rewrite
(\re{duhest}) as
\beqn\la{duhestr}
\!\!\!\!\!\!\!\!\!\!\!\!\!\!\!\!\!\!&&\!\!\!\!\!\! \Vert
\bP_1Z_1(t)\Vert_{-2-\de}\le\fr{C(\ov d_1)} {(1+t)^{1+\de}}\Vert
Z(0)\Vert_{\beta}
\nonumber\\
\!\!\!\!\!\!\!\!\!\!\!\!\!\!\!\!\!\!&&\!\!\!\!\!\! +\,C(\ov
d_1)\int_0^t\fr1{(1+|t-s|)^{1+\de}}\left[\Vert Z(s)\Vert_{-\beta}
\int_s^{t_1}\Vert Z(\tau)\Vert^2_{-\beta}d\tau+ \Vert
Z(s)\Vert^2_{-\beta}\right]ds,\,\,\,0\le t\le t_1, \eeqn
Let us
introduce the {\it majorant} \be\la{maj} m(t):=
\sup_{s\in[0,t]}(1+s)^{1+\de}\Vert
Z(s)\Vert_{-\beta}~,~~~~~~~~~t\in [0,t_*). \ee
To estimate $d_{1}(t)$ by $m(t_1)$ we note that
\be\la{wen}
w(s)-v(t_{1})=w(s)-v(s)+v(s)-v(t_{1})=
\dot c(s)+\int_s^{t_1}\dot v(\tau)d\tau
\ee
by (\re{vw}).
Hence, (\ref {dd1}),
Lemma \ref{mod} and the definition (\re{maj}) imply
\begin{eqnarray}
\!\!\!\!\!\! \!\!\! \!\!\!
 |d_{1}(t)|\!\!\! &=&\!\!\!|\int_{t_{1}}^{t}(w(s)-v(t_{1}))ds|\leq
\int_{t}^{t_{1}}\left( |\dot{c}(s)|+\int_{s}^{t_{1}}|\dot{v}(\tau )|d\tau
\right)ds   \nonumber  \label{d1est} \\
&&  \nonumber \\
\!\!\! &\leq &\!\!\!Cm^{2}(t_{1})\int_{t}^{t_{1}}\left( \frac{1}{(1+s)^{2+2\de}}%
+\int_{s}^{t_{1}}\frac{d\tau }{(1+\tau )^{2+2\de}}\right) ds\leq
Cm^{2}(t_{1}),~~~~0\leq t\le t_{1}.
\end{eqnarray}
We can
replace in (\re{duhestr}) the constants $C(\ov d_1)$ by $C$ if $m(t_1)$ is
bounded for $t_1\ge 0$. In order to do this replacement, we reduce
the exit time.
 Let us denote by
$\ve$ a fixed positive number which we will specify below.
\begin{definition} $t_{*}'$ is the exit time
\be\la{t*'}
t_*'=\sup \{t\in[0,t_*):
m(s)\le \ve,~~0\le s\le t\}.
\ee
\end{definition}
Now (\re{duhestr}) and (\re{d1est}) imply that for $t_1<t_*'$ \beqn\la{duhestri}
\!\!\!\!\!\!\!\!\!\!\!\!\!\!\!\!\!\!&&\!\!\!\!\!\! \Vert
\bP_1Z_1(t)\Vert_{-2-\de}\le\fr{C} {(1+t)^{1+\de}}\Vert
Z(0)\Vert_{\beta}
\nonumber\\
\!\!\!\!\!\!\!\!\!\!\!\!\!\!\!\!\!\!&&\!\!\!\!\!\!
+C\int_0^t\fr1{(1+|t-s|)^{1+\de}}\left[\Vert Z(s)\Vert_{-\beta}
\int_s^{t_1}\Vert Z(\tau)\Vert^2_{-\beta}d\tau+ \Vert
Z(s)\Vert^2_{-\beta}\right]ds,\,\,\,0\le t\le t_1. \eeqn


\setcounter{equation}{0}

\section{Symplectic Orthogonality}

Finally, we are going to change $\bP_1Z_1(t)$ by $Z(t)$ in the
left hand side of (\re{duhestri}). We will prove that it is
possible using again that $d_\beta\ll 1$ in (\re{close}) and due to
the following important bound:

\begin{lemma}\la{Z1P1Z1}
For sufficiently small $\ve>0$, we have for $t_1<t_*'$:
\be\la{Z1P1est}
\Vert Z(t)\Vert_{-2-\de}\le C\Vert \bP_1Z_1(t)\Vert_{-2-\de},
~~~~0\le t \le t_1,
\ee
where $C$ depends only on $\rho$ and $\ti v$.
\end{lemma}
{\bf Proof }
The proof is based on
the symplectic orthogonality
(\re{PZ}), i.e.
\be\la{PZ1}
\bPi_{v(t)}Z(t)=0,~~~~t\in[0,t_1],
\ee
and on the fact
that
 all the spaces $\cZ(t):=\bP_{v(t)}\cE$ are almost parallel
for all $t$.


Namely, we first note that
$\Vert Z(t)\Vert_{-2-\de}\le C(\ve)\Vert Z_1(t)\Vert_{-2-\de}$
by Lemma \re{dest}, since $|d_1(t)|\le C\ve^2$ for
 $t\le t_1<t_*'$ by (\re{d1est}).
Therefore, it suffices
 to prove that
\be\la{Z1P1ests}
\Vert Z_1(t)\Vert_{-2-\de}\le 2\Vert \bP_1Z_1(t)\Vert_{-2-\de},
~~~~~~~~0\le t\le t_1.
\ee
This estimate will follow from
\be\la{Z1P1estf}
\Vert\bPi_{v_1}Z_1(t)\Vert_{-2-\de}\le\fr12\Vert Z_1(t)\Vert_{-2-\de},
\,\,\,0\le t\le t_1\,,
\ee
since $\bP_1Z_1(t)=Z_1(t)-\bPi_{v_1}Z_1(t)$, where $v_1=v(t_1)$.
To prove (\re{Z1P1estf}), we write (\re{PZ1}) as
\be\la{PZ0r}
\bPi_{v(t),1}Z_1(t)=0,~~~~t\in[0,t_1],
\ee
where $\bPi_{v(t),1}Z_1(t)$ is $\bPi_{v(t)}Z(t)$ expressed in
terms of the variable $y_1=y+d_1(t)$.
Hence, (\re{Z1P1estf}) follows from  (\re{PZ0r}) if
the difference
 $\bPi_{v_1}-\bPi_{v(t),1}$ is small uniformly in $t$,
i.e.
\be\la{difs}
\Vert(\bPi_{v_1}-\bPi_{v(t),1})Z_1(t)\Vert_{-2-\de}<\fr12\,\Vert Z_1(t)\Vert_{-2-\de},~~~~~~~
0\le t\le t_1.
\ee
It remains to justify (\ref{difs}) for a sufficiently small
$\varepsilon
>0.$
In order to prove the bound (\re{difs}), we will need the formula
(\ref{Piv}) and the following relation which follows from
(\ref{Piv}):
\begin{equation}
\mbox{\boldmath $\Pi$}_{v(t),1}Z_{1}(t)=\sum \mbox{\boldmath $\Pi$}%
_{jl}(v(t))\tau _{j,1}(v(t))\Omega (\tau _{l,1}(v(t)),Z_{1}(t)),  \label{P1}
\end{equation}
where $\tau _{j,1}(v(t))$ are the vectors $\tau _{j}(v(t))$ expressed in the
variables $y_{1}$. In detail (cf. (\ref{inb})),
\begin{equation}
\left.
\begin{array}{rcl}
\tau _{j,1}(v) & := & (-\partial _{j}\psi _{v}(y_{1}-d_{1}(t)),-\partial
_{j}\pi _{v}(y_{1}-d_{1}(t)),e_{j},0), \\
\tau _{j+3,1}(v) & := & (\partial _{v_{j}}\psi _{v}(y_{1}-d_{1}(t)),\partial
_{v_{j}}\pi _{v}(y_{1}-d_{1}(t)),0,\partial _{v_{j}}p_{v}),
\end{array}
\right| ~~~j=1,2,3,  \label{inb*}
\end{equation}
where $v=v(t)$. Thus, we have to estimate the difference of
$$
{\bf \Pi}_{v_1}Z_1(t)=\sum{\bf \Pi}_{jl}(v_1)\tau_j(v_1,y_1)\Om(\tau_l(v_1,y_1),Z_1(t,y_1))
$$
and
$$
{\bf \Pi}_{v(t),1}Z_1(t)=\sum{\bf \Pi}_{jl}(v(t))\tau_j(v(t),y_1-d_1(t))
\Om(\tau_l(v(t),y_1-d_1(t)),Z_1(t,y_1)).
$$
The estimate is based on the following bounds.
First,
\begin{equation}
|\mbox{\boldmath $\Pi$}_{jl}(v(t))-\mbox{\boldmath
$\Pi$}_{jl}(v(t_{1}))|= |\int_{t}^{t_{1}}\dot{v}(s)\cdot \nabla
_{v}\mbox{\boldmath $\Pi$}_{jl}(v(s))ds|\leq
C\int_{t}^{t_{1}}|\dot{v}(s)|ds,~~~~0\leq t\leq t_{1}, \label{013}
\end{equation}
since $|\nabla_{v}\mbox{\boldmath $\Pi$}_{jl}(v(s))|$ is uniformly
bounded by (\re{sit}).
Second,
$$
|\Om(\tau_l(v_1,y_1)-\tau_l(v(t),y_1-d_1(t)),Z_1(t,y_1))|\le\Vert\tau_l(v_1,y_1)
-\tau_l(v(t),y_1-d_1(t))\Vert_{2+\de}\Vert Z_1(t,y_1)\Vert_{-2-\de}.
$$
Further, since $|d_{1}(t)|\leq C\ve^2$ and $\nabla \tau _{j}$ are smooth and sufficiently fast
decaying at infinity functions, Lemma \re{dest}
implies
\begin{equation}
\Vert \tau _{j,1}(v(t))-\tau _{j}(v(t))\Vert_{2+\de}\leq
C|d_{1}(t)|\le C\ve^{2},~~~~0\leq t\leq t_{1}  \label{011}
\end{equation}
for all $j=1,2,\dots,6$, where $C$ depends only on $\de$ and $\ti v$.
Finally,
\[
\tau _{j}(v(t))-\tau _{j}(v(t_{1}))=\int_{t}^{t_{1}}\dot{v}(s)\cdot \nabla
_{v}\tau _{j}(v(s))ds,
\]
and therefore
\begin{equation}
\Vert \tau _{j}(v(t))-\tau _{j}(v(t_{1}))\Vert_{2+\de}\leq
C\int_{t}^{t_{1}}|\dot{v}(s)|ds,~~~~0\leq t\leq t_{1}.  \label{012}
\end{equation}
At last, the bounds (\ref{difs}) will follow
from (\ref{Piv}), (\ref{P1}) and (\ref{011})-(\ref{013}) if we
establish that the integral in the right hand
side of (\ref{012}) and (\re{013}) can be made as small as we please by choosing
$\varepsilon >0$ sufficiently small.
Indeed,
\begin{equation}
\int_{t}^{t_{1}}|\dot{v}(s)|ds\leq Cm^{2}(t_{1})\int_{t}^{t_{1}}\frac{ds}{%
(1+s)^{2+2\de}}\leq C\varepsilon ^{2},~~~~0\leq t\le t_{1}.
\label{tvjest}
\end{equation}
The proof is complete.\hfill
$\Box$


\setcounter{equation}{0}

\section{Decay of Transversal Component}
Here we complete the proof of Proposition \re{pdec}.

\medskip

\noindent {\it Step i)} We fix an $\ve$, $0<\ve\le r_{-\beta}(\ti v)$ and $t'_*=t'_*(\ve)$
for which Lemma \re{Z1P1Z1} holds.
Then the bound of type (\re{duhestri})
holds with
$\Vert \bP_1Z_1(t)\Vert_{-2-\de}$ in the left hand side replaced by
 $\Vert Z(t)\Vert_{-\beta}$~:
\beqn\la{duhestrih}
\!\!\!\!\!\!\!\!\!\!\!\!\!\!\!\!\!\!&&\!\!\!\!\!\! \Vert
Z(t)\Vert_{-\beta}\le \Vert Z(t)\Vert_{-2-\de}\le C\Vert \bP_1Z_1(t)\Vert_{-2-\de}\le\fr{C}
{(1+t)^{1+\de}}\Vert Z(0)\Vert_{\beta}
\nonumber\\
\!\!\!\!\!\!\!\!\!\!\!\!\!\!\!\!\!\!&&\!\!\!\!\!\!
+C\int_0^t\fr1{(1+|t-s|)^{1+\de}}\left[\Vert Z(s)\Vert_{-\beta}
\int_s^{t_1}\Vert Z(\tau)\Vert^2_{-\beta}d\tau+ \Vert
Z(s)\Vert^2_{-\beta}\right]ds,\,\,\,0\le t\le t_1 \eeqn for
$t_1<t_*'$. This implies an integral inequality
 for the majorant $m(t)$ introduced by (\re{maj}).
Namely, multiplying both sides of (\re{duhestrih}) by
$(1+t)^{1+\de}$, and taking the supremum in $t\in[0,t_1]$, we get
$$
m(t_1) \le C\Vert Z(0)\Vert_{\beta}+ C\sup_{t\in[0,t_1]}\ds
\int_0^t\fr{(1+t)^{1+\de}}{(1+|t-s|)^{1+\de}}\left[\fr{m(s)}{(1+s)^{1+\de}}
\int_s^{t_1}\fr{m^2(\tau)d\tau}{(1+\tau)^{2+2\de}}+\fr{m^2(s)}
{(1+s)^{2+2\de}}\right]ds
$$
for $t_1\le t_*'$. Taking into account that $m(t)$ is a monotone
increasing function, we get
\be\la{mest}
m(t_1)\le C\Vert Z(0)\Vert_{\beta}+C[m^3(t_1)+m^2(t_1)]I(t_1),
~~~~~~~~~~~~~t_1\le t_*'.
\ee
where
$$
I(t_1)= \sup_{t\in[0,t_1]}
\int_0^{t}\fr{(1+t)^{1+\de}}{(1+|t-s|)^{1+\de}}\left[\fr1{(1+s)^{1+\de}}
\int_s^{t_1}\fr{d\tau}{(1+\tau)^{2+2\de}}+\fr1{(1+s)^{2+2\de}}\right]ds
\le \ov I<\infty,~~~~t_1\ge0.
$$
Therefore, (\re{mest}) becomes
\be\la{m1est}
m(t_1)\le C\Vert Z(0)\Vert_{\beta}+C\ov I[m^3(t_1)+m^2(t_1)],~~~~ t_1<t_*'.
\ee
This inequality implies that $m(t_1)$
is bounded for $t_1<t_*'$, and moreover,
\be\la{m2est}
m(t_1)\le C_1\Vert Z(0)\Vert_{\beta},~~~~~~~~~t_1<t_*'\,,
\ee
since $m(0)=\Vert Z(0)\Vert_{\beta}$ is sufficiently small by
 (\re{closeZ}).

\medskip

\noindent {\it Step ii)} The constant $C_1$ in the estimate
(\re{m2est}) does not depend on
$t_*$ and $t_*'$ by Lemma \re{Z1P1Z1}.
We choose $d_{\beta}$ in (\re{close}) so small that
$\Vert Z(0)\Vert_{\beta}<\ve/(2C_1)$. It is possible due to (\re{closeZ}).
Then the estimate (\re{m2est}) implies that $t'_*=t_*$ and therefore
 (\re{m2est}) holds for all $t_1<t_*$.
Then the bound (\re{d1est}) holds for all $t<t_*$.
Therefore,
(\re{Zt}) also holds for all $t<t_*$.
Finally, this implies that
$t_*=\infty$, hence also $t'_*=\infty$ and
(\re{m2est}) holds for all $t_1>0$ if $d_{\beta}$ is small enough.

The transversal decay (\re{Zdec}) is proved.\hfill $\Box$

\section{Soliton Asymptotics}
\setcounter{equation}{0}

Here we prove our main
Theorem \re{main} relying on
the transversal decay (\re{Zdec}). First we will prove the asymptotics
(\re{qq}) for the vector components, and afterwards
the asymptotics (\re{S}) for the fields.

\subsection
{Asymptotics for the vector components}
From (\re{addeq}) we
have $\dot q=\dot b+\dot Q$, and from (\re{reduced}), (\re{redN}) with $\beta=4+\de$, and
(\re{AA}) it follows that $\dot Q=B_{v(t)}P+{\cal O} (\Vert
Z\Vert^2_{-\beta})$. Thus,
\be\la{dq} \dot q=\dot b+\dot
Q=v(t)+\dot c(t)+B_{v(t)}P(t)+{\cal O} (\Vert Z\Vert^2_{-\beta}).
\ee The equation
(\re{parameq}) and the estimates (\re{NZ}), (\re{Zdec}) imply
\be\la{bv} |\dot c(t)|+|\dot v(t)|\le \ds\fr {C_1(\rho,\ov
v,d_\beta)}{(1+t)^{2+2\de}}, ~~~~~~t\ge0. \ee Therefore, $c(t)=c_+
+\cO(t^{-(1+2\de)})$ and $v(t)=v_+ +\cO(t^{-(1+2\de)})$,
$t\to\infty$. Since $|P|\le\Vert Z\Vert_{-\beta}$, the estimate
(\re{Zdec}), and (\re{bv}), (\re{dq}) imply that \be\la{qbQ} \dot
q(t)=v_++\cO(t^{-1-\de}). \ee Similarly, \be\la{bt}
b(t)=c(t)+\ds\int_0^tv(s)ds=v_+t+a_++\cO(t^{-2\de}), \ee hence the
second part of (\re{qq}) follows: \be\la{qbQ2}
q(t)=b(t)+Q(t)=v_+t+a_++\cO(t^{-2\de}), \ee since
$Q(t)=\cO(t^{-1-\de})$ by  (\re{Zdec}).

\bigskip

\subsection{Asymptotics for the fields}
We apply the approach
developed in \ci{IKSs}, see also \ci{IKM,IKSm,IKSr,KKS}.
For the field part of the solution, $F(t)=(\psi(x,t),\pi(x,t))$
let us define the {\it accompanying soliton field} as
$$
F_{\rm v(t)}(t)=(\psi_{\rm v(t)}(x-q(t)),\pi_{\rm v(t)}(x-q(t))),
$$
where we define now ${\rm v}(t)=\dot q(t)$, cf. (\re{dq}).
Then for the difference $Z(t)=F(t)-F_{\rm v(t)}(t)$ we obtain
easily the equation \ci{KKS}, Eq. (2.5),
$$
\dot Z(t)=AZ(t)-\dot{\rm v}\cdot\na_{\rm v}F_{{\rm v}(t)}(t),\,\,\,\,\,\,A(\psi,\pi)=(\pi,\De\psi).
$$
Then \be\la{eqacc} Z(t)=W_0(t)Z(0)-\int_0^tW_0(t-s)[\dot{\rm
v}(s)\cdot\na_{\rm v}F_{{\rm v}(s)}(s)]ds. \ee
Since $\Vert(\psi_{v_+},\pi_{v_+})(x-v_+t-a_+)-F_{{\rm v}(t)}(t)\Vert_{\cal F}
={\cal O}(t^{-2\de})$ by (\re{qbQ}) and (\re{qbQ2}),
to obtain the
asymptotics (\re{S}) it suffices to prove that
$Z(t)=W_0(t)\Psi_++r_+(t)$ with some $\Psi_+\in{\cal F}$ and
$\Vert r_+(t)\Vert_{{\cal F}}={\cal O}(t^{-\de})$. This is
equivalent to \be\la{Sme} W_0(-t)Z(t)=\bPsi_++r_+'(t), \ee where
$\Vert r_+'(t)\Vert_\cF=\cO(t^{-\de})$ since $W_0(t)$ is a unitary
group in the Sobolev space $\cF$ by the energy conservation for
the free wave equation. Finally,
 (\re{Sme}) holds since
 (\re{eqacc}) implies that
\be\la{duhs} W_0(-t)Z(t)= Z(0)+\int_0^t
W_0(-s)R(s)ds,\,\,\,\,\,R(s)=\dot{\rm v}(s)\cdot\na_{\rm v}F_{{\rm
v}(s)}(s), \ee where the integral  in the right hand side of
(\re{duhs}) converges in the Hilbert space $\cF$ with the rate
$\cO(t^{-\de})$. The latter holds since $\Vert
W_0(-s)R(s)\Vert_\cF =\cO(s^{-1-\de})$ by the unitarity of
$W_0(-s)$ and the decay rate $\Vert R(s)\Vert_\cF
=\cO(s^{-1-\de})$.
Let us prove this rate of decay. It suffices to prove that
$|\dot {\rm v}(s)|=\cO(s^{-1-\de})$, or equivalently
$|\dot p(s)|=\cO(s^{-1-\de})$. Substitute (\re{add}) to the
last equation of (\re{system0}) and obtain \beqn \dot p(t)
&=&\int\left[\psi_{v(t)}(x-b(t))+\Psi(x-b(t),t)\right] \na\rho(x-b(t)-Q(t))dx
\bigskip\\ &=&\int\psi_{v(t)}(y)\na\rho(y)dy+
\int\psi_{v(t)}(y)\left[\na\rho(y-Q(t))-\na\rho(y)\right]dy +\int\Psi(y,t)\na\rho(y-Q(t))dy.
\nonumber
\eeqn
The first integral in the right hand side is zero by the stationary
equations (\re{stfch}). The second integral is ${\cal O}(t^{-1-\de})$,
since $Q(t)={\cal O}(t^{-1-\de})$, and by the conditions (\re{ro}) on $\rho$.
Finally, the third integral is ${\cal O}(t^{-1-\de})$ by the estimate (\re{Zdec}).
The proof is complete.\hfill $\Box$


\appendix


\setcounter{equation}{0}

\section{Structure of the matrix $M^{-1}(i\om)$}

We prove Lemmas \re{162} and \re{Minv}. Recall that for $\om\in\R$
$$
M(i\om)=\left(
\ba{cc}
i\om I_3  & -B_{v} \\
-F(i\om) & i\om I_3 \ea \right),
$$
where
$$
B_v=\left(
\ba{ccc}
\nu^3 & 0 & 0\\
0 & \nu & 0\\
0 & 0 & \nu
\ea
\right),\,\,\,F(i\om)=\left(
\ba{ccc}
f_{1}(\om) & 0 & 0\\
0 & f(\om) & 0\\
0 & 0 & f(\om)
\ea
\right).
$$
Here $f_1(\om)=F_{11}(i\om+0),\,\,f(\om)=F_{22}(i\om+0)=F_{33}(i\om+0)$ with
\be\la{Fjjlam} F_{jj}(\lam)=\int dk\fr{|\hat\rho|^2k_j^2}{k^2+(\lam+ik_1v)^2}-\int
dk\fr{|\hat\rho|^2k_j^2}{k^2-(k_1v)^2}. \ee  $F_{jj}(\lam)$ are
analytic functions in $\C$ by Lemma \re{cmf}. Thus,
$$
F_{jj}(\lam)=F_{jj}(0)+F'_{jj}(0)\lam+\fr{F''_{jj}(0)}{2}\lam^2+\dots
$$
Here $F_{jj}(0)=0$ by (\re{Fjjlam}). Further, by (\re{Fjjlam}) \be\la{Fpjj}
F'_{jj}(\lam)=-2\int dk\,k_j^2|\hat\rho|^2\fr{\lam+ivk_1}{(k^2+(\lam+ivk_1)^2)^2} \ee
and
$$
F'_{jj}(0)=-2iv\int dk\,k_j^2|\hat\rho|^2\fr{k_1}{(k^2-(vk_1)^2)^2}=0,
$$
since the integrand function is odd in $k_1$. Hence, we obtain
$F_{jj}(\lam)=\lam^2r_j(\lam)$, where $r_j(\lam)$ is analytic in $\C$. Note that
$r_j(0)=F''_{jj}(0)/2$. By (\re{Fpjj}) we have
$$
F''_{jj}(\lam)=-2\int dk\,k_j^2|\hat\rho|^2\fr{k^2-3(\lam+ivk_1)^2}
{(k^2+(\lam+ivk_1)^2)^3}
$$
and finally, \be\la{Ij0} F_{jj}(i\om)=-\om^2r_j(\om),\,\,\,\, r_j(0)=-\int
dk\,k_j^2|\hat\rho|^2\fr{k^2+3(vk_1)^2}{(k^2-(vk_1)^2)^3}. \ee Let us denote
$r(\om)=r_{2}(\om)=r_{3}(\om)$. Then
\be\la{invinv}
L(\om):=M^{-1}(i\om)=\left(
\ba{cc}
L_{11}(\om) & L_{12}(\om) \\
L_{21}(\om) & L_{22}(\om)
\ea
\right),
\ee
where
\be\la{inv11}
L_{11}(\om)=\left(
\ba{ccc}
\fr{\ds -i\om}{\ds \om^2+\nu^3f_1(\om)} & 0 & 0 \\
0 & \fr{\ds -i\om}{\ds \om^2+\nu f(\om)} & 0 \\
0 & 0 & \fr{\ds -i\om}{\ds \om^2+\nu f(\om)} \ea \right)=\fr1{\om}\left( \ba{ccc}
\fr{\ds -i}{\ds 1-\nu^3r_1(\om)} & 0 & 0 \\
0 & \fr{\ds -i}{\ds 1-\nu r(\om)} & 0 \\
0 & 0 & \fr{\ds -i}{\ds 1-\nu r(\om)} \ea \right)
\ee
by (\re{Ij0}); we denote the last matrix ${\cal L}_{11}(\om)$. Similarly,
\be\la{inv12}
L_{12}(\om)=\left(
\ba{ccc}
\fr{\ds -\nu^3}{\ds \om^2+\nu^3f_1(\om)} & 0 & 0 \\
0 & \fr{\ds-\nu}{\ds\om^2+\nu f(\om)} & 0 \\
0 & 0 & \fr{\ds-\nu}{\ds\om^2+\nu f(\om)} \ea \right)= \fr1{\om^2}\left( \ba{ccc}
\fr{\ds-\nu^3}{\ds 1-\nu^3r_1(\om)} & 0 & 0 \\
0 & \fr{\ds-\nu}{\ds 1-\nu r(\om)} & 0 \\
0 & 0 & \fr{\ds-\nu}{\ds 1-\nu r(\om)} \ea \right),
\ee
we denote the last matrix ${\cal L}_{12}(\om)$. Note that
$$
{\cal L}_{11}(\om)=i{\cal L}_{12}(\om)B_v^{-1}.
$$
Further,
\be\la{inv21}
L_{21}=\left(
\ba{ccc}
\fr{\ds-f_1(\om)}{\ds \om^2+\nu^3f_1(\om)} & 0 & 0 \\
0 & \fr{\ds-f(\om)}{\ds \om^2+\nu f(\om)} & 0 \\
0 & 0 & \fr{\ds-f(\om)}{\ds \om^2+\nu f(\om)} \ea \right)= \left( \ba{ccc}
\fr{\ds r_1(\om)}{\ds 1-\nu^3r_1(\om)} & 0 & 0 \\
0 & \fr{\ds r(\om)}{\ds 1-\nu r(\om)} & 0 \\
0 & 0 & \fr{\ds r(\om)}{\ds 1-\nu r(\om)} \ea \right),
\ee
so we put ${\cal L}_{21}=L_{21}$. Finally,
\be\la{inv22}
L_{22}(\om)=L_{11}(\om)=\fr1{\om}{\cal L}_{11}(\om),
\ee
and thus, ${\cal L}_{22}(\om)={\cal L}_{11}(\om)$. Note that the denominators of the matrix
elements of each matrix ${\cal L}_{11}$ to ${\cal L}_{22}$ are nonzero at $\om=0$,
since $r_1(0)<0$ and $r(0)<0$ by (\re{Ij0}). For $\om\ne0$ the denominators are nonzero
 by Lemma \re{lW}. This completes
the proof of Lemma \re{Minv}. Finally, (\re{invinv}) to (\re{inv22}) imply Lemma \re{162},
 since $r_j(\om)=-F_{jj}(i\om)/\om^2$, where $F_{jj}(i\om)$ are bounded functions by (\re{Flambound}).


\setcounter{equation}{0}

\section{Symplectic orthogonality conditions}

Let us check that the symplectic orthogonality conditions
(\re{ort16}) with
$j=1,2,3$ read the first equation of (\re{ort123}). By (\re{phidef}),
$$
\Phi(\Psi_0,\Pi_0)(0)=i\langle ikv\hat\Psi_0+\hat\Pi_0,
\fr{k\hat\rho}{\hat D(0)}
\rangle,\,\,\,~~~
\hat D(0)=k^2-(kv)^2.
$$
On the other hand,
by (\re{ort16})  with
$j=1,2,3$, and (\re{inb}), (\re{solF}), we have
$$
0=\Om(X_0,\tau_j)=-\langle\Psi_0,\pa_j\pi_v\rangle+\langle\Pi_0,\pa_j\psi_v\rangle-P_0\cdot e_j=
\langle
\hat\Psi_0,\fr{k_j kv\hat\rho}{\hat D(0)}\rangle
+\langle\hat\Pi_0,
\fr{ik_j\hat\rho}{\hat D(0)}
\rangle
-P_0\cdot e_j
$$
$$
~~~~=
\langle kv \hat\Psi_0,
\fr{ k_j\hat\rho}{\hat D(0)}
\rangle
-i
\langle\hat\Pi_0,
\fr{k_j\hat\rho}
{\hat D(0)}\rangle
-(P_0)_j=-\Phi_j(\Psi_0,\Pi_0)(0)-(P_0)_j.
$$

\bigskip

\noindent
Now let us check that the conditions
(\re{ort16}) with
$j=4,5,6$
read the second equation of (\re{ort123}). By (\re{phidef})
$$
\Phi'(0)=
i\langle
\fr{\hat\Psi_0}{\hat D(0)},k\hat\rho\rangle
-i\langle
2ikv\fr{ikv\hat\Psi_0+\hat\Pi_0}{\hat D^2(0)},
k\hat\rho
\rangle
=\langle
\fr{(k^2+(kv)^2)i\hat\Psi_0+2kv\Pi_0}{(k^2-(kv)^2)^2},k\hat\rho
\rangle,
$$
where
the integral converges by the condition (\re{zero4}).
On the other hand,
by (\re{ort16})  with
$j=4,5,6$, and (\re{inb}), (\re{solF}), we have
for
$j=1,2,3$
$$
0=\Om(X_0,\tau_{3+j})=
\langle \hat\Psi_0, \pa_{v_j}\pi_v\rangle
-
\langle \hat\Pi_0, \pa_{v_j}\psi_v\rangle
+
Q_0\cdot\pa_{v_j}p_v
$$
$$
=\langle
\hat\Psi_0,
-ik_j\fr{k^2+(kv)^2}{\hat D^2(0)}
\hat\rho
\rangle
+\langle\hat\Pi_0,2k_j
\fr{kv\hat\rho}{\hat D^2(0)}
\rangle
+Q_0\cdot\pa_{v_j}p_v
=\Phi'_j(0)+(B_v^{-1}Q_0)_j,
$$
since $Q_0\cdot\pa_{v_j}p_v=Q_0\cdot B_v^{-1}e_j=B_v^{-1}Q_0\cdot e_j$.


{\bf Email adresses of the authors:}\\

ivm61@mail.ru

\bigskip

alexander.komech@univie.ac.at

\bigskip

brvainbe@uncc.edu


\begin{thebibliography}{99}

 \bibitem{A}
M. Abraham, Theorie der Elektrizitat, Band 2: Elektromagnetische Theorie der
Strahlung, Teubner, Leipzig (1905).

\bibitem{BC08} D. Bambusi and S. Cuccagna, {\it On dispersion of small energy solutions of
the nonlinear Klein Gordon equation with a potential},
http://arxiv.org/abs/0908.4548.

\bibitem{BL} H. Beresticky and P.L. Lions,
 Nonlinear scalar field equations,
{\em  Arch. Rat. Mech. and Anal. }  {\bf 82} (1983),
no.4, 313-375.



\bibitem{BP1} V.S. Buslaev and G.S. Perelman, {\it On nonlinear scattering of
states which are close to a soliton},  in ``M\'{e}thodes
Semi-Classiques, Vol.2 Colloque International (Nantes, juin 1991)'',
Asterisque, {\bf 208} (1992) 49-63.

\bibitem{BP2} V.S. Buslaev and G.S. Perelman, {\it Scattering for the nonlinear
Schr\"odinger equation: states close to a soliton}, St.
Petersburg Math. J., {\bf 4} (1993), 1111-1142.

\bibitem{BP3} V.S. Buslaev and G.S. Perelman, {\it On the stability of solitary waves
for nonlinear Schr\"odinger equations}, Nonlinear evolution equations (N.N.Uraltseva,
eds.), Transl. Ser. 2, 164, Amer. Math. Soc., Providence, RI, 1995, pp. 75-98.


\bibitem{BS} V.S. Buslaev, C. Sulem,  {\it
On asymptotic stability of solitary waves for nonlinear Schr\"odinger
equations},  Ann. Inst. Henri Poincar\'e, Anal. Non Lin\'eaire, {\bf  20}  (2003), No.3, 419-475.

\bibitem{Cu} S. Cuccagna, {\it Stabilization of solutions to nonlinear
Schr\"odinger equations}, Comm. Pure Appl. Math., {\bf 54} (2001), 1110-1145.

\bibitem{Cu03} S. Cuccagna, {\it On asymptotic stability of ground states of NLS},
Rev. Math. Phys., {\bf 15} (2003), 877-903.

\bibitem{Cu10} S. Cuccagna, {\it The Hamiltonian structure of the nonlinear
Schr\"odinger equation and the asymptotic stability of its ground states}, http://
arxiv.org/abs/0910.3797.

\bibitem{CM08} S. Cuccagna and T. Mizumachi, {\it On asymptotic stability in energy space
of ground states for Nonlinear Schr\"odinger equations}, Comm. Math. Phys., {\bf 284}
(2008), 51-87.

\bibitem{EH81} W. Eckhaus and A. van Harten,
``The Inverse Scattering Transformation and the Theory of Solitons.
An Introduction'', North-Holland, Amsterdam, 1981.

\bibitem{EGS}
M. Esteban, V. Georgiev, and E. Sere,
 Stationary solutions of the Maxwell-Dirac
and the Klein-Gordon-Dirac equations,
{\em Calc. Var. Partial Differ. Equ.}
{\bf  4} (1996), no.3,
265-281.


\bibitem{FZ93}
 A.S. Fokas, V.E. Zakharov,
Important Developments in Soliton Theory,
Springer, Berlin, 1993.




\bibitem{IKM} V. Imaikin, A. Komech, and P. Markowich, {\it Scattering of
solitons of the Klein-Gordon equation coupled to a classical particle},
J. Math. Phys., {\bf 44} (2003), 1202-1217.


\bibitem{IKSm} V. Imaikin, A. Komech, and H. Spohn,
{\it Soliton-like asymptotics and scattering for a particle coupled to
 Maxwell field}, Russ. J. Math. Phys., {\bf 9} (2002), 428-436.

\bibitem{IKSs} V. Imaikin, A. Komech, and H. Spohn,
{\it Scattering theory for a particle coupled to a scalar field},
Discrete Contin. Dyn. Syst., {\bf 10} (2003), 387-396.

\bibitem{IKSr} V. Imaikin, A. Komech, and H. Spohn,
{\it Rotating charge coupled to the Maxwell field:
scattering theory and adiabatic limit}, Monatsh. Math., {\bf 142} (2004),
143-156.

\bibitem{IKV05} V. Imaikin, A. Komech, and B. Vainberg,
{\it On scattering of solitons for the Klein-Gordon equation coupled to a particle},
Comm. Math. Phys., {\bf 268} (2006), 321-367.

\bibitem{IKV06} V. Imaikin, A. Komech, and B. Vainberg,
{\it On scattering of solitons for wave equation coupled to
 a particle}, pp. 249-273 in:
``CRM Proceedings and Lecture Notes'', {\bf 42} (2007).

\bibitem{KKS} A. Komech, M. Kunze, and H. Spohn, {\it Effective dynamics for a
mechanical particle coupled to a wave field}, Comm. Math. Phys.,
{\bf 203} (1999), 1-19.

\bibitem{KSK}  A. Komech, M. Kunze, and H. Spohn, {\it Long-time asymptotics for a
classical particle interacting with a scalar wave field}, Comm. Partial
Differential Equations, {\bf 22} (1997), 307-335.

\bibitem{KS} A.I. Komech and H. Spohn, {\it Soliton-like asymptotics for a
classical particle interacting with a scalar wave field}, Nonlinear Anal., {\bf 33} (1998), 13-24.

 \bibitem{KS00} A.I. Komech and H. Spohn,
Long-time asymptotics for the coupled Maxwell-Lorentz equations,
{\em Comm. Partial Diff. Eqns.} {\bf 25} (2000), no. 3/4, 558-585.


\bibitem{IKM04} V. Imaikin, A.I. Komech and N. Mauser.
Soliton-type asymptotics for the coupled Maxwell-Lorentz equations,
{\em Ann. Inst. Poincar\'e, Phys. Theor.}
{\bf 5} (2004),
1117-1135.

\bibitem{Kop} E.A. Kopylova, {\it Weighted energy decay for 3D wave equation},
Asymptotic Anal., {\bf 65} (2009), 1-16.

\bibitem{MM08} Y. Martel and F. Merle, {\it Asymptotic stability of solitons of the gKdV
equations with general nonlinearity}, Math. Ann., {\bf 341} (2008), 391-427.

\bibitem{MW96} J. Miller and M. Weinstein, {\it Asymptotic stability of solitary
waves for the regularized long-wave equation}, Comm. Pure Appl. Math., {\bf  49}  (1996), 399-441.

\bibitem{PW92} R.L. Pego and M.I. Weinstein, {\it On asymptotic stability of solitary
waves}, Phys. Lett. A, {\bf 162} (1992), 263-268.

\bibitem{PW94} R.L. Pego and M.I. Weinstein, {\it Asymptotic stability of solitary waves},
Comm. Math. Phys., {\bf 164} (1994), 305-349.

\bibitem{RS4} M. Reed and B. Simon, ``Methods of modern mathematical physics.
Vol. IV, Analysis of operators'', Acad. Press, N.-Y., 1978.

\bibitem{Sig93} I.M. Sigal, {\it Nonlinear wave and Schr\"odinger equations. I.
Instability of periodic and quasiperiodic solutions}, Comm. Math. Phys., {\bf 153}
(1993), 297-320.

\bibitem{SW88} A. Soffer and M.I. Weinstein, {\it Multichannel nonlinear scattering
for nonintegrable systems}, in ``Proceedings of Conference on an
Integrable and Nonintegrable Systems, June, 1988, Oleron, France,
Integrable Systems and Applications'', Springer Lecture Notes in
Physics, Volume 342.

\bibitem{SW90} A. Soffer and M.I. Weinstein, {\it Multichannel nonlinear scattering in
nonintegrable systems}, Comm. Math. Phys., {\bf 133} (1990),
119-146.

\bibitem{SW92} A. Soffer and M.I. Weinstein, {\it Multichannel nonlinear scattering and
stability II. The case of Anisotropic and potential and data},
J. Differential Equations, {\bf 98} (1992), 376-390.

\bibitem{SW3} A. Soffer and M.I. Weinstein, {\it Resonances, radiation damping
and instability in Hamiltonian nonlinear wave equations}, Invent. Math., {\bf 136} (1999), 9-74.

\bibitem{SW4} A. Soffer and M.I. Weinstein, {\it Selection of the ground state
for nonlinear Schr\"odinger equations}, ArXiv:nlin.PS/0308020, 2003.

\bibitem{Sp} H. Spohn, ``Dynamics of Charged Particles and Their Radiation Field'',
Cambridge University Press, Cambridge, 2004.

\bibitem{Vai} B. Vainberg, ``Asymptotic methods in equations of mathematical physics'',
Gordon and Breach Publishers, New-York-London, 1989.


\end{thebibliography}
\end{document}